\documentclass[aps,prl,twocolumn,amsmath,amssymb,nofootinbib,superscriptaddress]{revtex4-2}
\usepackage[colorlinks,linkcolor=blue,anchorcolor=blue,allcolors = blue,citecolor=blue]{hyperref}
\usepackage{bm}
\usepackage{braket}
\usepackage{graphicx}
\usepackage{enumerate}
\usepackage{enumitem}
\usepackage{makecell}
\usepackage{xcolor}
\usepackage{ulem}
\usepackage{mathtools}
\usepackage{booktabs}
\usepackage{subfig}

\newcommand{\affa}{Beijing Key Laboratory of Fault-Tolerant Quantum Computing, \\Beijing Academy of Quantum Information Sciences, Beijing 100193, China}
\newcommand{\affb}{Beijing National Laboratory for Condensed Matter Physics, \\Institute of Physics, Chinese Academy of Sciences, Beijing 100190, China}
\newcommand{\affc}{School of Physical Sciences, University of Chinese Academy of Sciences, Beijing 100190, China}
\newcommand{\affd}{Hefei National Laboratory, Hefei 230088, China}
\newcommand{\affe}{Center for Joint Quantum Studies, Department of Physics, School of Science, Tianjin University, Tianjin 300072, China}
\newcommand{\afff}{College of Computer Science and Technology, National University of Defense Technology, Changsha 410073, China}
\begin{document}

\title{Realization of Quantum Switch with Superconducting Qubits}

\author{Z. H. Yang}
\thanks{These authors contributed equally to the work.}
\address{\affa}
\address{\affb}
\address{\affc}
\address{\afff}
\author{Z. T. Wang}
\thanks{These authors contributed equally to the work.}
\address{\affa}
\address{\affe}
\author{Kaixuan Huang}
\thanks{These authors contributed equally to the work.}
\address{\affa}
\author{Jingning Zhang}
\email{zhangjn@baqis.ac.cn}
\address{\affa}
\author{\mbox{H. K. Xu}}
\address{\affa}
\author{Hao Li}
\address{\affa}
\author{Kui Zhao}
\address{\affa}
\author{Wei-Guo Ma}
\address{\affb}
\author{S. K. Zhao}
\address{\affa}
\author{Yulong Feng}
\address{\affa}
\author{Jiaxiu Han}
\address{\affa}
\author{Guangming Xue}
\address{\affa}
\author{Kai Xu}
\address{\affb}
\address{\affa}
\address{\affd}
\author{Haifeng Yu}
\email{hfyu@baqis.ac.cn}
\address{\affa}
\address{\affd}
\author{Heng Fan}
\email{hfan@iphy.ac.cn}
\address{\affb}
\address{\affa}
\address{\affd}
\author{S. P. Zhao}
\email{spzhao@iphy.ac.cn}
\address{\affa}
\address{\affb}

\pacs{xxx}

\begin{abstract}
We report the experimental realization of quantum switch with superconducting qubits, which employs a logical qubit encoding scheme that extends the Hilbert space containing both vacuum and logical basis states within the two-qubit computational space. We experimentally emulate the quantum communication processes through noisy channels and show the enhanced communication achieved by employing quantum switch. Clear enhancement in quantum communication capacity is also experimentally demonstrated. Importantly, we show that our approach can be used for solving the promise problems in the general \(N\)-switch scenario and serve as a scalable platform for the study of quantum computational advantage. 
\end{abstract}

\maketitle

{\it Introduction}---In the quantum world, a system can be in the superposition of different quantum states. Likewise, the orders of different events or operations can also be placed in a quantum superposition, in contrast to the usual fixed orders in the classical world. Such indefinite causal order (ICO)~\cite{PhysRevA.86.040301, oreshkov2012quantum, PhysRevA.88.022318} has received considerable interest over the past years, being actively studied both theoretically and experimentally~\cite{Rozema2024, deng2025generalized}.

\begin{figure}[t]
	\centering
	\includegraphics[width=0.35\textwidth]{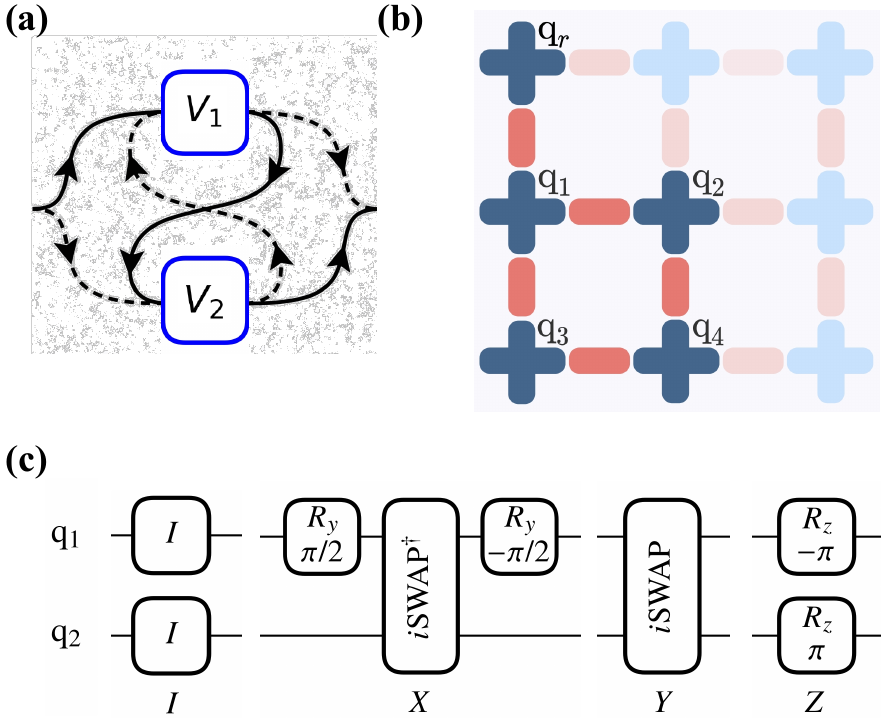}
	\caption{(a) Superposition of opposite orders of $V_1$ and $V_2$ operations realized by QS. (b) Schematic of a 3$\times$3 transmon qubit lattice with twelve tunable couplers. Five qubits and couplers (dark crosses and rectangles) are used in the experiment. (c) Circuits for logical qubit Pauli operators using physical qubits $q_1$ and $q_2$.}
	\label{fig1}
\end{figure}

Quantum switch (QS) is the best studied ICO process, which has found a number of applications in quantum communication or communication complexity~\cite{buhrman2010nonlocality, PhysRevLett.122.120504, PhysRevLett.117.100502, PhysRevLett.120.120502, procopio2019communication, PhysRevResearch.2.033292, PhysRevLett.124.030502}, quantum computation (e.g. promise problems)~\cite{PhysRevLett.113.250402, procopio2015experimental, PRXQuantum.2.010320, PhysRevResearch.3.043012, PhysRevLett.128.230503}, quantum metrology~\cite{frey2019indefinite, PhysRevLett.124.190503, PhysRevA.103.032615, PhysRevLett.131.090801, yin2023experimental}, thermodynamics~\cite{PhysRevLett.125.070603, PhysRevA.102.032215, PhysRevA.105.032217}, and quantum reference frames~\cite{bnkn-4p3f}. The basic process is illustrated in Fig.~\ref{fig1}(a) in the simplest 2-switch scenario, in which \(V_{1}\) and \(V_{2}\) can be different communication channels, quantum gates, position and momentum displacements, and heat or workflow, depending on the specific applications. The solid line indicates the order for the first implementation of \(V_{1}\) followed by that of \(V_{2}\), while the dashed line indicates the opposite order. Numerous studies show that by using the superposition of the two opposite orders, the advantages in various applications can be achieved~\cite{buhrman2010nonlocality, PhysRevLett.122.120504, PhysRevLett.117.100502, PhysRevLett.120.120502, procopio2019communication, PhysRevResearch.2.033292, PhysRevLett.124.030502, PhysRevLett.113.250402, procopio2015experimental, PRXQuantum.2.010320, PhysRevResearch.3.043012, PhysRevLett.128.230503, frey2019indefinite, PhysRevLett.124.190503, PhysRevA.103.032615, PhysRevLett.131.090801, yin2023experimental, PhysRevLett.125.070603, PhysRevA.102.032215, PhysRevA.105.032217}.

So far the experimental studies have been overwhelmingly based on the photonic platform~\cite{PhysRevLett.122.120504, PhysRevResearch.2.033292, PhysRevLett.124.030502, procopio2015experimental, PRXQuantum.2.010320, yin2023experimental}. In this Letter, we report the first experimental realization of QS using superconducting qubits, which employs logical qubit encoding where the vacuum state and logical computational space are embedded in the Hilbert space of two physical qubits. To show the full functionality of the ICO process, we experimentally emulate the 2-switch quantum communication processes through noisy channels and obtain the high-fidelity or noiseless quantum information transmission. Significant enhancement in quantum capacity compared to the case without ICO is also experimentally demonstrated. 

Superconducting qubits are a natural choice for implementing more general \(N\)-switch due to their scalability and programmability. We show that our approach can be used for solving the Hadamard promise problems in the 4-switch scenario, which were proposed for experimentally accessible implementation of QS beyond $N$=2~\cite{PRXQuantum.2.010320}. This is done by numerical simulation using state-of-the-art experimental parameters to circumvent the present device limitation, which clearly shows the feasibility of superconducting qubits as a scalable platform for the study of quantum computational advantage using ICO.

{\it Experiment and logical qubit encoding}---In our experiment, we used five qubits from a 3$\times$3 transmon qubit lattice with 12 tunable couplers, as shown schematically in Fig.~\ref{fig1}(b) (see Supplemental Material~\cite{supplementary} for device parameters and gate performance). In photonic experiments~\cite{Rozema2024}, the control and target registers are encoded in the path and polarization degrees of freedom, respectively. The Hilbert space of each optical path is spanned by the vacuum state \(|\mathrm{vac}\rangle\) and the polarization-encoded qubit states \(|H\rangle \coloneqq |0\rangle\) and \(|V\rangle \coloneqq |1\rangle\). Crucially, \(|\mathrm{vac}\rangle\) remains unchanged when the polarization is manipulated by wave plates \cite{Quantum2023}. For superconducting qubits, an independent vacuum degree of freedom can be assigned if we have an expanded Hilbert space. This can be achieved by logical qubit encoding on two physical qubits. Namely, the basis states for two physical qubits are \(\{|00\rangle ,|01\rangle ,|10\rangle ,|11\rangle \}\), from which we choose three states to encode the vacuum state and the two basis states of the logical qubit~\cite{remark}:
\begin{equation}
|\mathrm{vac}\rangle \coloneqq |00\rangle,\quad |0\rangle_L  \coloneqq |01\rangle,\quad |1\rangle_L  \coloneqq |10\rangle~.\label{eq:logical_states}
\end{equation}
The logical Pauli operators \(I, X, Y\), and \(Z\), as well as their rotations, are defined with respect to the basis \(\{|0\rangle_L, |1\rangle_L \}\) and leave the vacuum invariant~\cite{supplementary}. The circuits for logical Pauli operators using physical qubit rotational and $i$SWAP gates are exemplified in Fig.~\ref{fig1}(c).

{\it Emulation of quantum communication}---The use of QS to enhance communication rates through noisy channels was proposed theoretically~\cite{PhysRevLett.120.120502} and later experimentally demonstrated in a photonic system~\cite{PhysRevLett.124.030502}. To emulate the process with superconducting qubits, we consider two Pauli channels:
${\mathcal E}(\rho) = \sum_{i = 0}^{3}p_{i}\sigma_{i}\rho \sigma_{i},~{\mathcal F}(\rho) = \sum_{i = 0}^{3}q_{i}\sigma_{i}\rho \sigma_{i}$,
where $p_i$ and $q_i$ $(i=0, 1, 2, 3)$ are probability vectors satisfying $\sum_{i} p_{i} = \sum_{i} q_{i}= 1$, and \(\sigma_{i} \in \{I, X, Y, Z\}\) are the Pauli operators with \(\sigma_0 = I\). The two different orders ${\mathcal {EF}}$ and ${\mathcal {FE}}$ correspond to the control qubit states $|x\rangle_c$ = $|0\rangle_c$ and $|1\rangle_c$, respectively. It can be shown that on the control qubit basis $\ket{\pm}_c = \left(\ket{0}_c\pm i\ket{1}_c\right)/\sqrt{2}$, the QS can be written as~\cite{supplementary}
 \begin{equation}
\begin{aligned}
    {\mathcal S}({\mathcal E}, {\mathcal F})(\ket{+}\bra{+}_{c} \otimes \rho) & =  \ket{+}\bra{+}_{c} \otimes P_{+} {\mathcal C}_{+}(\rho)\\  & + \ket{-}\bra{-}_{c} \otimes P_{-} {\mathcal C}_{-}(\rho)~,\\
	\mathcal{C}_{\pm}(\rho) = \frac{1}{4P_{\pm}}\sum_{i,j = 0}^{3} & p_{i}q_{j}[\sigma_{i},\sigma_{j}]_{\pm}\rho [\sigma_{i},\sigma_{j}]_{\pm}^{\dagger}~,
\end{aligned}
\label{commun}
\end{equation}
where \([A, B]_{\pm} = AB \pm BA\). $P_{-}$ = $(1-p_0)(1-q_0)-\sum_{i=1}^3p_iq_i$ and $P_{+}=1-P_{-}$ are the post-selection probabilities. QS applies operations in different orders given by the control qubit, and the final state is projectively measured on the $|\pm \rangle_c$ basis. The channels \(\mathcal{C}_{\pm}\) determine how effectively quantum information is transmitted.

\begin{figure}[t]
	\centering
	\includegraphics[width=0.45\textwidth]{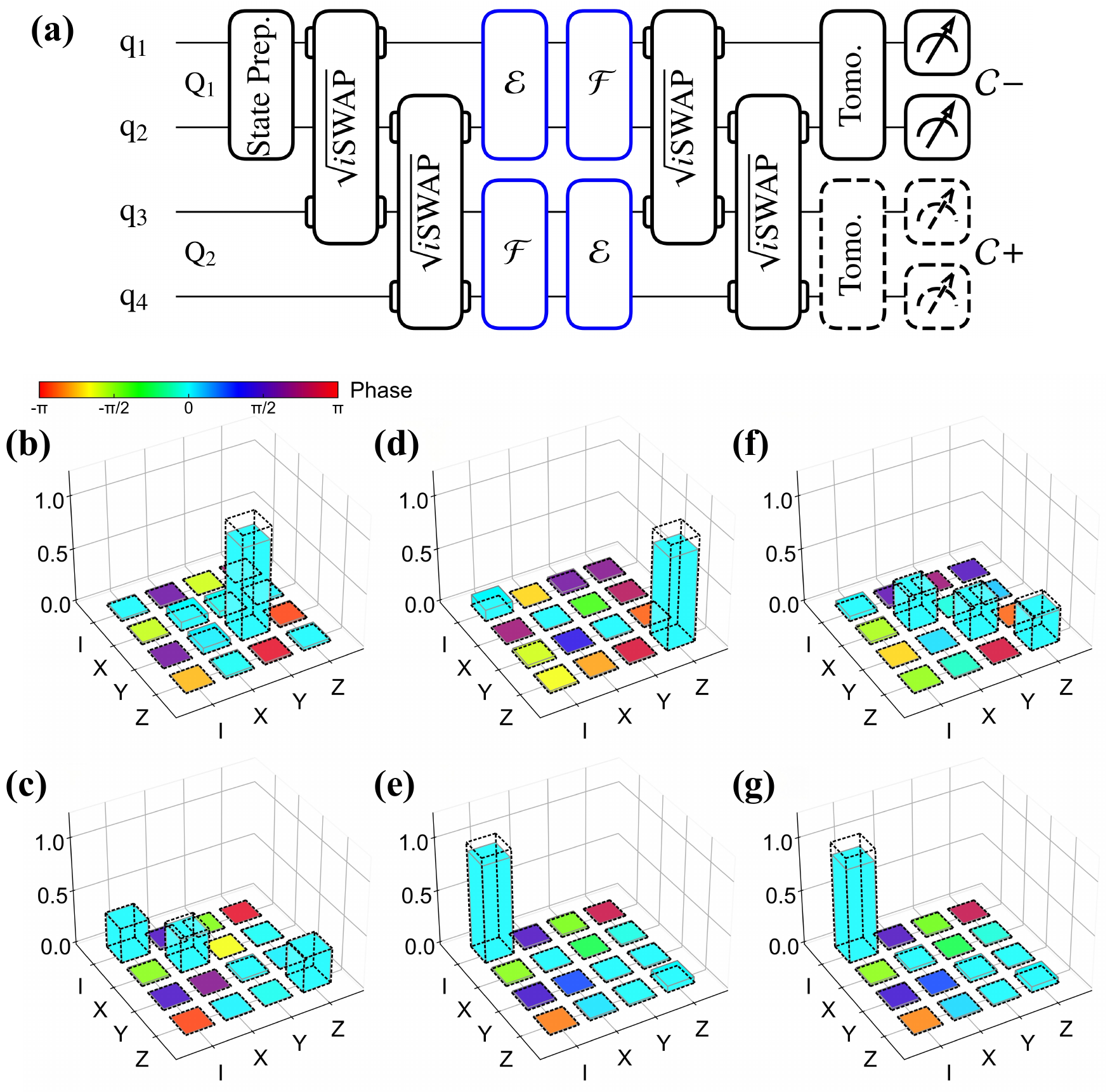}
	\caption{(a) Circuit to emulate quantum communication using two logical qubits $Q_1$ and $Q_2$. Experimentally reconstructed matrices of $\mathcal{C}_{-}$ and $\mathcal{C}_{+}$ for (b, c) ${\mathcal S}({\mathcal P}_1, {\mathcal P}'_1)$ with $s=t=0.5$, (d, e) ${\mathcal S}({\mathcal P}_2, {\mathcal P}_2)$ with $p_1=p_2=0.5$, and (f, g) ${\mathcal S}({\mathcal P}_3, {\mathcal P}_3)$ with $p_1=p_2=p_3=1/3$. Post selection is used to mitigate decoherence and gate infidelity. The black dashed lines show the theoretical results from Eq.~(\ref{commun}) in the three cases: $\mathcal{C}_{-}(\rho)$ = $\sigma_2\rho\sigma_2$, $\sigma_3\rho\sigma_3$, and ($\sigma_1\rho\sigma_1$ + $\sigma_2\rho\sigma_2$ + $\sigma_3\rho\sigma_3$)/3; $\mathcal{C}_{+}(\rho)$ = ($\sigma_0\rho\sigma_0$ + $\sigma_1\rho\sigma_1$ + $\sigma_3\rho\sigma_3$)/3, $\sigma_0\rho\sigma_0$, and $\sigma_0\rho\sigma_0$.}
	\label{fig2}
\end{figure}

Figure~\ref{fig2}(a) shows the circuit to implement the QS using superconducting logical qubits. The physical qubits $(q_1, q_2)$ and $(q_3, q_4)$ form the two logical qubits $Q_1$ and $Q_2$. The logical qubit $Q_1$ is first prepared in the $|0\rangle_L$, $|1\rangle_L$, $(|0\rangle_L+|1\rangle_L)/\sqrt{2}$, $(|0\rangle_L-|1\rangle_L)/\sqrt{2}$, $(|0\rangle_L+i|1\rangle_L)/\sqrt{2}$, and $(|0\rangle_L-i|1\rangle_L)/\sqrt{2}$ states. Then, two parallel $\sqrt{i\rm SWAP}$ gates, acting as a 50:50 beam splitter in photonic experiment~\cite{PhysRevLett.124.030502}, are applied, followed by ${\mathcal E}$ and ${\mathcal F}$ operations in opposite orders. After mapping the control qubit to the $|\pm\rangle_c$ basis via $\sqrt{i\rm SWAP}$ gates, quantum process tomography is performed on $Q_1$ and $Q_2$ to obtain the reconstructed matrix $\gamma_{ij}$ in the channel $\mathcal{C}_{\pm}(\rho)$ = $\sum_{ij}\gamma_{ij}\sigma_{i}\rho\sigma_{j}$.

\begin{figure*}[t]
	\centering
	\includegraphics[width=0.9\textwidth]{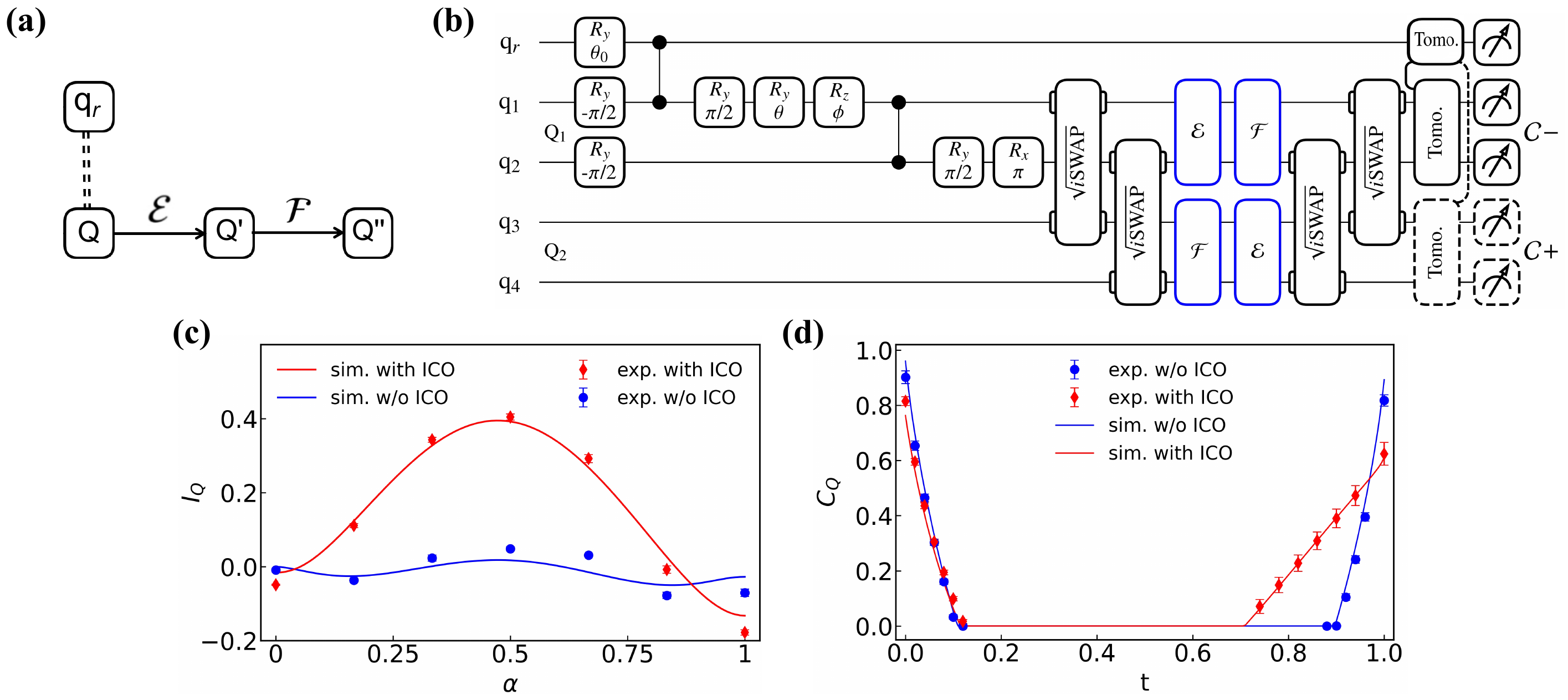}
	\caption{Coherent information $I_Q$ and quantum capacity $C_Q$. (a) Schematic of using reference qubit $q_r$ to purify the initial state $\rho$. (b) Circuit for measuring coherent information $I_Q$. (c) $I_Q$ as a function of the initial-state parameter $\alpha$ for channel $\mathcal{S}({\mathcal P}_1, {\mathcal P}'_1)$ with $s = t = 0.9$ and $\theta = \phi = 0$. Blue symbols and lines represent experimental results and numerical simulations without ICO, respectively, while red symbols and lines show the corresponding data with ICO. (d) $C_Q$  as a function of parameters $t = s$ for $\mathcal{S}({\mathcal P}_1, {\mathcal P}'_1)$ with $\alpha = 0.5$ and $\theta = \phi = 0$. Blue symbols (lines) represent the experimental results (simulations) without ICO, while those in red show the corresponding data with ICO. Post selection is used to mitigate decoherence and gate infidelity. Error bars represent the standard deviation from 200 and 30 experimental repetitions in (c) and (d), respectively, each having 6000 shots.}
	\label{fig3}
\end{figure*}

We consider three QS channels $\mathcal{S}({\mathcal P}_1, {\mathcal P}'_1)$, $\mathcal{S}({\mathcal P}_2, {\mathcal P}_2)$, and $\mathcal{S}({\mathcal P}_3, {\mathcal P}_3)$ with Pauli channels (1) ${\mathcal P}_1(\rho) = (1-s)\rho + s\sigma_1\rho\sigma_1$, (2) ${\mathcal P}'_1(\rho) = (1-t)\rho + t\sigma_3\rho\sigma_3$, (3) ${\mathcal P}_2(\rho) = \sum_{i=1}^{2}p_i\sigma_i\rho\sigma_i$, and (4) ${\mathcal P}_3(\rho) = \sum_{i=1}^{3}p_i\sigma_i\rho\sigma_i$. The experimental results of the reconstructed matrices $\gamma_{ij}$ for $\mathcal{S}({\mathcal P}_1, {\mathcal P}'_1)$ at $s=t=0.5$,  $\mathcal{S}({\mathcal P}_2, {\mathcal P}_2)$ at $p_1 = p_2 = 0.5$, and $\mathcal{S}({\mathcal P}_3, {\mathcal P}_3)$ at $p_1 = p_2 = p_3 = 1/3$ are shown in Figs.~\ref{fig2}(b, c), (d, e), and (f, g), respectively (see Ref.~\cite{supplementary} for detailed circuit implementation). The dashed lines are the theoretical results from Eq.~(\ref{commun}). The fidelity,  defined by $F(\chi_{\rm exp},\chi_{\rm ideal})={\rm tr}(\chi^{\dagger}_{\rm ideal}\chi_{\rm exp})$, where $\chi_{\rm exp}$ and $\chi_{\rm ideal}$ are the experimental and ideal process matrices~\cite{PhysRevA.71.062310, PhysRevLett.112.050502}, is found to be $F$=$90.9\%$ ($98\%$), $F$=$93.5\%$ ($89.9\%$), and $F$=$91.6\%$ ($96.7\%$) for channel $\mathcal{C}_{-}$ ($\mathcal{C}_{+}$), respectively.

The channel ${\mathcal P}_1(\rho)$ [${\mathcal P}'_1(\rho)$] is known as the bit-flip (phase-flip) channel~\cite{nielsen2010}, which keeps the $x$-axis ($z$-axis) projection of the Bloch vector while the components along other directions are gradually lost as $s$ (or $t$) increases from 0 toward 0.5. At $s$ (or $t$) = 0.5, the Bloch vector contracts completely to the $x$-axis ($z$-axis). As a result, both combined channels ${\mathcal P}_1{\mathcal P}'_1$ and ${\mathcal P}'_1{\mathcal P}_1$ will not allow for any quantum information transmission. The channel ${\mathcal P}_2(\rho)$ behaves in a similar way as ${\mathcal P}'_1(\rho)$, with the Bloch vector contracting in the $xy$ directions but with different rates depending on the specific values of $p_1$ and $p_2$. At $p_1 = p_2 = 0.5$, the Bloch vector shrinks to the $z$ axis. For ${\mathcal P}_3(\rho)$, the Bloch vector reduces in all directions with different rates for different $p_1, p_2, p_3$ values. At $p_1 = p_2 = p_3 = 1/3$, the Bloch sphere shrinks to a reduced sphere of radius 1/3~\cite{supplementary}.  Therefore, the experimental results in Fig.~\ref{fig2} demonstrate that for noisy channels which partially or completely block the transmission of quantum information, the use of QS may allow for a noiseless transmission. For example, Figs.~\ref{fig2}(b) and (d) show the unitary gates $\sigma_2$ and $\sigma_3$ for the $\mathcal{C}_{-}(\rho)$ channel, while Figs.~\ref{fig2}(e) and (g) show the identity gate $\sigma_0$ for the $\mathcal{C}_{+}(\rho)$ channel. The transmitted information thus can be accurately decoded through these channels.

{\it Quantum capacity}---To experimentally quantify the channel capacity~\cite{PhysRevA.55.1613, IEEE2005} of transmitting quantum information, we use a reference qubit $q_r$, which couples to the logical qubit $Q_1$ to purify the initial state $\rho$~\cite{schumacher_pra1996quantum}, as shown in Fig.~\ref{fig3}(a). The quantum capacity $C_Q$ is the maximum coherent information $I_Q$, defined as~\cite{nielsen2010, schumacher_pra1996quantum}: 
\begin{equation}
	I_Q = S_{vN}\bigl[{\mathcal E}(\rho)\bigr] - S_{vN}\bigl[{\mathcal I}_r \otimes {\mathcal E}(\ket{\Phi_\rho}\bra{\Phi_\rho})\bigr]~.
	\label{IQ}
\end{equation}
Here $S_{vN}(\rho) = -\mathrm{Tr}(\rho \log_2 \rho)$ is the von Neumann entropy, ${\mathcal I}_r$ is the identity on $q_r$, and $\ket{\Phi_\rho}$ is an arbitrary purification of $\rho$, which can be written as $\ket{\Phi_\rho} = \sqrt{\alpha}\ket{0}_r \otimes \ket{\psi} + \sqrt{1-\alpha}\ket{1}_r \otimes \ket{\psi_\perp}$ for the system state $\rho = \alpha \ket{\psi}\bra{\psi} + (1-\alpha )\ket{\psi_\perp}\bra{\psi_\perp}$, with $\{\ket{\psi}, \ket{\psi_\perp}\}$ being the orthonormal basis of the system.

The coherent information $I_Q$ measures the degree to which quantum coherence is retained after the state evolution from $\rho$ to ${\mathcal E}(\rho)$~\cite{schumacher_pra1996quantum}. On the right hand side of Eq.~(\ref{IQ}), the first term is the entropy of $\rho$ after evolution under ${\mathcal E}$, and the second term is that of $\ket{\Phi_\rho}\bra{\Phi_\rho}$ after evolution under ${\mathcal I}_r \otimes {\mathcal E}$. The circuit for measuring the second term is shown in Fig.~\ref{fig3}(b), while the first term can be obtained by tracing out $q_r$ from the measured results. In the figure, the purified state $\ket{\Phi_\rho}\bra{\Phi_\rho}$ is created by the CZ, $R_y(-\pi/2)$, and $R_y(\pi/2)$ gates. Subsequent channel operations are similar to those in Fig.~\ref{fig2}(a), but the final tomographic measurements involve $q_r$ for both $\mathcal{C}_{\pm}(\rho)$ channels and $I_Q=P_+I_Q(\mathcal{C}_+)+P_-I_Q(\mathcal{C}_-)$. 

In our experiment, we set $|\psi\rangle = \cos(\theta/2)|0\rangle_L + \sin(\theta/2) e^{i\phi}|1\rangle_L$, $|\psi_\perp\rangle = -\sin(\theta/2)|0\rangle_L + \cos(\theta/2) e^{i\phi}|1\rangle_L$ and prepare different states $\rho$ set by ($\alpha, \theta, \phi$)~\cite{PhysRevLett.124.030502}. The parameter $\alpha$ is controlled via a $R_y(\theta_0)$ gate giving $\alpha = \cos^2(\theta_0/2)$, and $\theta$ and $\phi$ are given by the $R_y(\theta)$ and $R_z(\phi)$ gates. The system undergoes the QS channel $\mathcal{S}({\mathcal P}_1, {\mathcal P}'_1)$, after which quantum state tomography is performed on the output states. The von Neumann entropies in Eq.~(\ref{IQ}) are obtained to calculate $I_Q$, and $C_Q$ is determined by maximizing $I_Q$ over input state $\rho$.

Figure~\ref{fig3}(c) shows the measured coherent information $I_Q$ as a function of $\alpha$ for $\mathcal{S}({\mathcal P}_1, {\mathcal P}'_1)$ with $s$ = $t$ = 0.9 and $\theta$ = $\phi$ = 0 (symbols). Here, $\alpha=0$, $0.5$, and $1$ correspond to the initial states of $\ket{1}_L$, the maximally mixed state, and $\ket{0}_L$, respectively. The ICO yields clear enhancement of $I_Q$, with a maximum of about $0.35$ near $\alpha=0.5$, whereas the data without ICO show nearly vanishing coherent information over the measured range. The solid lines are the simulation results, obtained using Qiskit~\cite{qiskit2024} and considering decoherence and the readout errors of $1\%$ for $\ket{0}_L$ and $2\%$ for $\ket{1}_L$. The experimental and simulation results agree well, including the small offsets from zero near $\alpha=0$ and $\alpha=1$, and the slight asymmetry of the $I_Q \sim \alpha$ curve with respect to $\alpha=0.5$. The quantum capacity $C_Q$ as a function of noise parameters $t = s$ for $\mathcal{S}({\mathcal P}_1, {\mathcal P}'_1)$ with $\alpha = 0.5$ and $\theta = \phi = 0$ is shown in Fig.~\ref{fig3}(d). Notably, in the range $0.72 < t < 0.95$, the quantum capacity with ICO significantly exceeds that without ICO, demonstrating again the advantage of using ICO for quantum information transmission.

{\it Hadamard promise problem}---QS provides advantages in some computational tasks like channel discrimination~\cite{PhysRevLett.113.250402, procopio2015experimental} or the Fourier~\cite{PhysRevLett.113.250402, PhysRevResearch.3.043012} and Hadamard~\cite{PRXQuantum.2.010320, PhysRevLett.128.230503} promise problems. In the Hadamard promise problem (HPP), $K$ unitary operations $\{U_j\}_{j=1}^{K}$ with $N\le K$! permutations are considered. The promise states that given an $N\times N$ Hadamard matrix $M_N$ with all +1 entries in the first row and column, the $x$th permutation $\Pi_x$ = $(M_N)_{x,y}\Pi_0$. The task is to find the right  permutation, or the column $y$ for the black-box unitaries. The algorithm for it is to first initialize the control-target system $\ket{0}_c\otimes\ket{\psi}$ and then apply the Hadamard transform $H_N=M_N/\sqrt{N}$ to the control register. Afterwards, the QS defined by $\mathcal{S}_N(|x\rangle_c\otimes |\psi\rangle) = |x\rangle_c\otimes \Pi_x|\psi \rangle$ and the inverse transform $H_N^{\dagger}$ on the control register are applied. If the Hadamard promise is satisfied, quantum interference singles out a unique column index $y$, yielding
\begin{equation}
	H_N^{\dagger}\mathcal{S}_N H_N(\ket{0}_c\otimes\ket{\psi}) = \ket{y}_c\otimes\Pi_0\ket{\psi},
\end{equation}
and $y$ can be determined by a single projective measurement of the control register~\cite{PRXQuantum.2.010320}. 

\begin{figure}[t]
	\centering
	\includegraphics[width=0.48\textwidth]{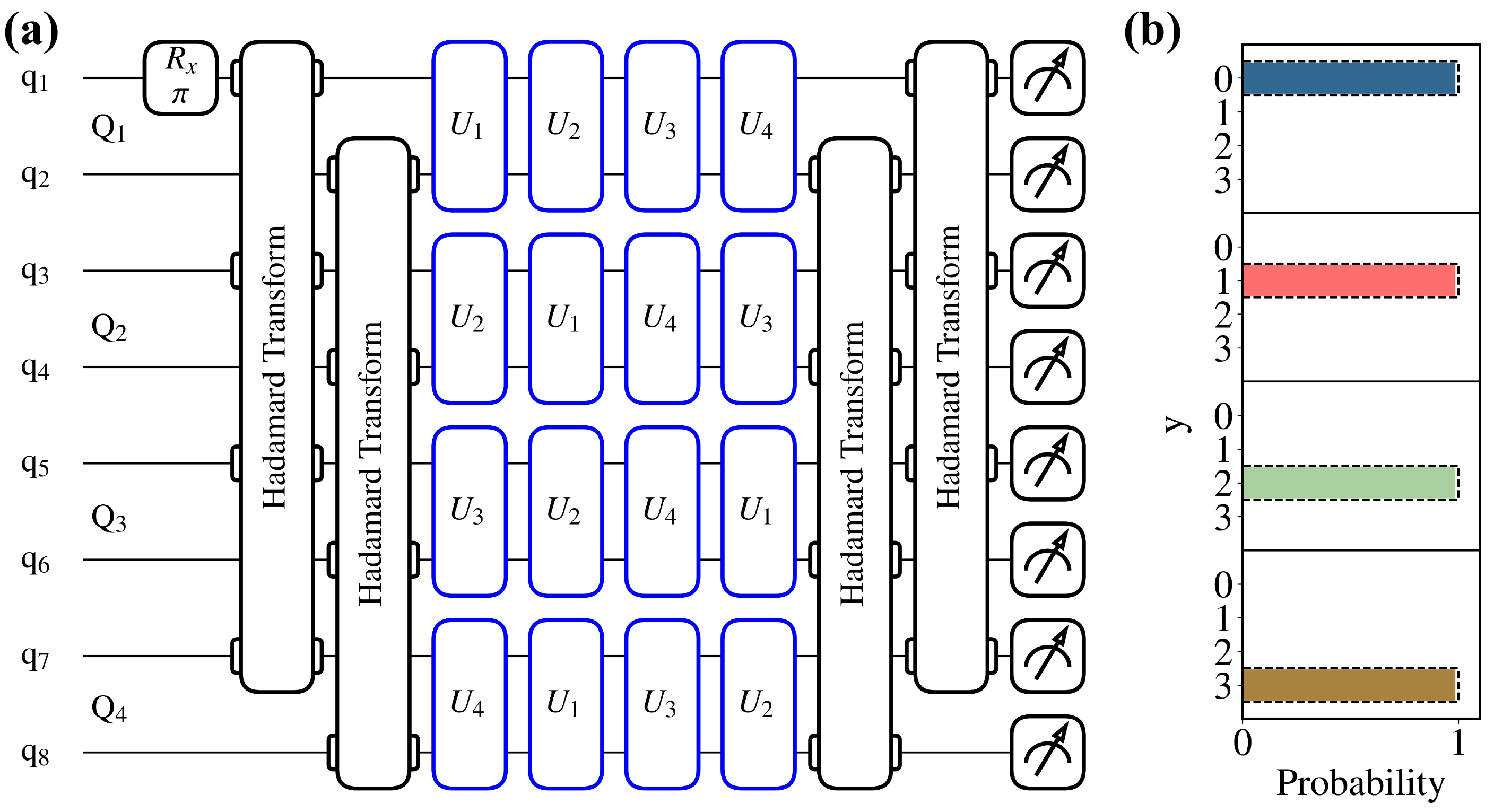}
	\caption{Solving an $N=K=4$ HPP on a $2\times 4$ ladder quantum processor. (a) Circuit with eight physical qubits ($q_1$--$q_8$) encoding four logical qubits ($Q_1$--$Q_4$), comprising an $X$ gate, Hadamard transforms, permuted unitary layers, and the readout. (b) Simulation results considering device decoherence for the gate sets in Table~\ref{tab:gate_sets}. Post selection is ued on the single-excitation subspace of the eight physical qubits. Dashed lines indicate the ideal results in the absence of decoherence.}
	\label{fig4}
\end{figure}

We demonstrate numerically that the algorithm can be implemented on a $2\times 4$ ladder processor for an HPP with $K=4$ unitary operations $\{U_1, U_2, U_3, U_4\}$ and $N=4$ permutations $\{1234, 2143, 3241, 4132\}$, with $\{1234\}$ representing $\Pi_0 = U_4 U_3 U_2 U_1$, etc~\cite{PRXQuantum.2.010320}. Each rung connects the two physical qubits encoding a logical qubit. The logical qubit $Q_1$ (the first rung), linked to the first permutation $\Pi_0$, is initialized by an excitation giving rise to the initial state $\ket{0}_c$. The single-excitation subspace of one rail corresponds to the $\ket{0}_L$ manifold and that of the other rail to the $\ket{1}_L$ manifold. The Hadamard transform in each rail can be implemented by a linear-depth circuit with single-qubit rotations and nearest-neighbor two-qubit gates. The circuit for the algorithm is shown in Fig.~\ref{fig4}(a), with the full circuit detailed in Ref.~\cite{supplementary}. 

\begin{table}[b]
	\centering
	\caption{Gate sets for an HPP with $N=K=4$.}
	\label{tab:gate_sets}
	\renewcommand{\arraystretch}{1.5}   
	\setlength{\tabcolsep}{10pt}        
	\begin{tabular}{ccccc}
		\hline \hline
		$y$ & $U_1$ & $U_2$ & $U_3$ & $U_4$ \\
		\midrule
		$0$ & $I$ & $X$ & $X$ & $X$ \\
		$1$ & $X$ & $Y$ & $Y$ & $Z$ \\
		$2$ & $X$ & $X$ & $Y$ & $Z$ \\
		$3$ & $X$ & $Y$ & $Z$ & $Z$ \\
		\hline \hline
	\end{tabular}
\end{table}

 Table~\ref{tab:gate_sets} lists the representative unitary sets that satisfy the Hadamard promise with respect to the columns of the Hadamard matrix
 	\begin{equation}
 		H_4 = \frac{1}{2}\left(\begin{array}{rrrr}
 			1 & 1 & 1 & 1\\
 			1 & -1 & 1 & -1\\
 			1 & 1 & -1 & -1\\
 			1 & -1 & -1 & 1
 		\end{array}\right),
 	\end{equation}
 	where the four promise classes are labeled by the column index $y$. To be more general, we consider continuum of gate sets and deliver each gate in an unknown basis, $U_i \mapsto W U_i W^\dagger$, with $W$ being a random $\mathrm{SU}(2)$ unitary~\cite{PRXQuantum.2.010320}. Since the promise is basis-independent, the conjugated sets satisfy it as well, and the ideal success probability of the quantum switch is unchanged. Figure~\ref{fig4}(b) displays the output $y$-distribution averaged over 20 random instances of $W$ for each promise class, simulated with the qubit coherence times $T_1 = 100\,\mu\mathrm{s}$ and $T_2 = 75\,\mu\mathrm{s}$, and gate duration of 30 ns for both single-qubit and CNOT gates taken from Ref.~\cite{Abanin2025}. The probability distributions are sharply peaked at distinct index $y$, with the probabilities of $0.991\pm0.001$, $0.9902\pm0.0008$, $0.989\pm0.001$, and $0.9889\pm0.0004$ for $y=0,1,2,3$, respectively. The fact that the quantum switch resolves $y$ in a single shot without any knowledge of $W$ demonstrates the feasibility of solving HPPs with random rotations on the superconducting processors.
 
 {\it Conclusion}---We have experimentally and numerically demonstrated the implementation of QS using logically encoded superconducting qubits. Experimental emulation of quantum information transmission through noisy channels was successfully performed, showing enhanced transmission and quantum capacity when ICO is employed. We further showed the feasibility of solving the general $N$-switch HPPs, thus demonstrating the superconducting qubits as a convenient and scalable platform for quantum computation utilizing ICO.

{\it Acknowledgments}---This work was supported by the National Natural Science Foundation of China (Grant Nos. 92365206, T2121001, 92265207, 12504593, T2322030, 92365301, 12504576 and 12404578), the Innovation Program for Quantum Science and Technology (Grant No. 2021ZD0301800), the Beijing Nova Program (Grant No. 20220484121), the Beijing National Laboratory for Condensed Matter Physics (Grant No. 2024BNLCMPKF022), and the China Postdoctoral Science Foundation (Grant No. GZB20240815).


\bibliography{ref}

\clearpage
\onecolumngrid

\appendix

\begin{center}
{\Large\bfseries Supplemental Material for}\\[0.6em]
{\large\bfseries ``Realization of Quantum Switch with Superconducting Qubits''}
\end{center}
\vspace{1.2em}

\renewcommand{\thefigure}{S\arabic{figure}}
\renewcommand{\thetable}{S\arabic{table}}
\renewcommand{\theequation}{S\arabic{equation}}
\setcounter{figure}{0}
\setcounter{table}{0}
\newcounter{appsec}
\renewcommand{\theappsec}{\Alph{appsec}}
\makeatletter
\let\old@section=\section
\renewcommand{\section}{\stepcounter{appsec}\setcounter{equation}{0}\old@section}
\makeatother
\renewcommand{\theequation}{\theappsec\arabic{equation}}

\section{\theappsec. Device information and measurement setup}

\begin{table}[b]
	\centering
	\caption{Basic parameters of q$_1$, q$_{2}$, q$_{3}$, q$_{4}$, and q$_{r}$. $\omega_r$, $\omega_{\rm{max}}$, and $\omega_{\rm{idle}}$ denote the readout resonator frequency, maximum qubit frequency and idle qubit frequency, respectively. $\alpha$ represents the qubit anharmonicity at the idle qubit frequency. $T_1$ denotes the energy-relaxation time, and $T_2^{*}$ denotes the Ramsey dephasing time, both measured at the idle point. $F_0$ and $F_1$ indicate readout fidelities for the ground ($|0\rangle$) and excited ($|1\rangle$) states. $F_{1q}$ denotes the single-qubit gate fidelity obtained from randomized benchmarking (RB). }
	\tabcolsep=0.9cm			
	\begin{tabular}{ccccccc}
		\hline \hline
		Qubit & q$_{1}$    &   q$_{2}$     &   q$_{3}$  & q$_{4}$    &   q$_{r}$      \\ \hline      
		$\omega_{r}/2\pi$(GHz) & 7.318   &   7.303     &   7.246   & 7.356   &   7.332   \\
		$\omega_{max}/2\pi$(GHz)      & 5.142    &   5.061     &   4.947  & 5.097  &   4.879      \\ 
		$\omega_{idle}/2\pi$(GHz) & 5.142    &   4.907    &   4.571  & 4.967    &   4.840     \\
		$\alpha/2\pi$(MHz)      & -198    &   -196     &   -196   & -196     &   -224  \\ 
		$T_{1}(\mu s)$ & 20.4    &   30.8     &   49.9   & 25.1    &   30.2      \\ 
		$T^{*}_{2}(\mu s)$      & 5.40   &   2.03    &   1.56  & 2.73    &   6.7    \\ 
		$F_0(\%)$      & 96.6    &   94.7     &   98.5  & 96.2    &   97.5     \\ 
		$F_1(\%)$      & 93.8   &   91.4     &  96.8  & 94.4    &   93.1    \\ \hline
		$F_{1q}(\%)$      & 99.951    &   99.962    &  99.849  & 99.952    &   99.940    \\	\hline	\hline	
	\end{tabular}	
	\label{tab:table1} 
\end{table}

\begin{figure}[t] 
    \centering 
    \includegraphics[width=0.99\textwidth]{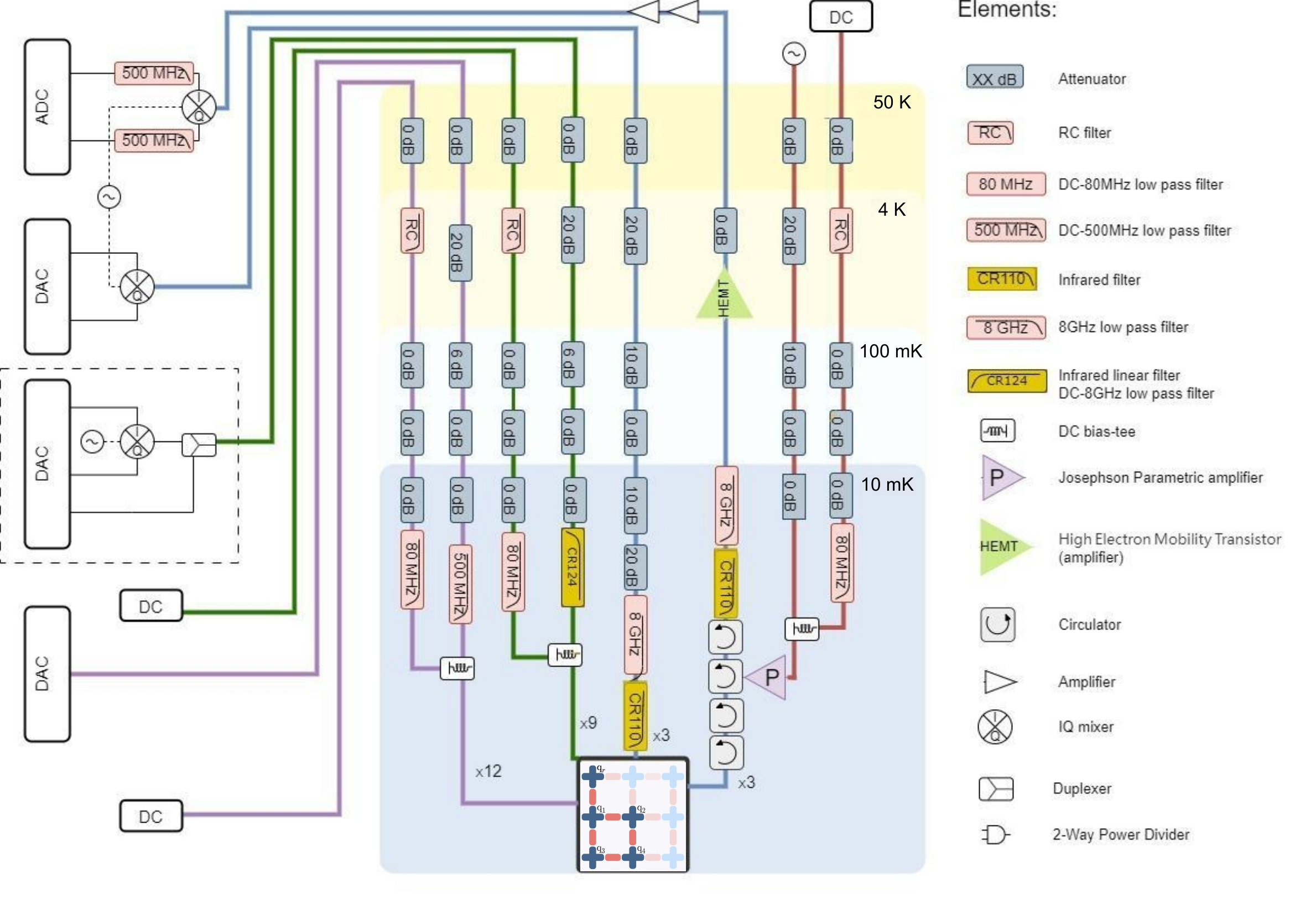}
    \caption{\centering Schematic of the measurement setup inside and outside the dilution refrigerator.}
    \label{fig:system} 
\end{figure}

The superconducting processor used for the emulation of quantum information transmission in this work consists of 9 transmon qubits and 12 couplers, fabricated using flip-chip technology with a tantalum base layer~\cite{li2021vacuum}. The device is arranged in a $3 \times 3$ square lattice, as shown in Fig. 1(b) in the main text and at the bottom of Fig.~\ref{fig:system}. In the schematic, cross symbols denote qubits, and rectangular bars represent couplers. Each qubit is equipped with independent control lines. During the experiment, the physical-qubit pairs $\rm (q_1,q_2)$ and $\rm (q_3,q_4)$ encode the logical qubits $Q_1$ and $Q_2$, respectively, while $\rm q_r$ serves as the reference qubit. These five active qubits are isolated from unused circuit elements by tuning the related couplers to their decoupling points. Their idle frequencies are selected through calibration to balance coherence time, residual coupling, and gate performance. Detailed device parameters, readout fidelity, and single-qubit gate performance are provided in Table~\ref{tab:table1}.

The measurement setup inside and outside the dilution refrigerator is shown in Fig.~\ref{fig:system}. For each qubit, the XY and Z control lines are first combined by a diplexer, then attenuated at the 50~K, 4~K, and 100~mK stages to suppress thermal photons from higher-temperature stages, filtered by a CR124 filter, and finally combined with the DC line at a bias tee before being delivered to the qubit. Couplers are controlled by combining only the Z control line with the DC line through a bias tee. The readout line is split into input and output paths. The input line passes through a series of attenuators and filters to provide a clean probe tone to the sample, while the output signal passes through four circulators, a Josephson parametric amplifier (JPA), a high-electron-mobility transistor (HEMT), and room-temperature microwave amplifiers before digitization. Each circulator provides approximately 15~dB of isolation, giving a nominal total isolation of 45~dB and thereby helping suppress noise propagating from the output chain back to the sample. For further details regarding the device parameters and single-qubit calibration procedures, the readers are referred to Ref.~\cite{wang_demonstration_2024}.

\section{\theappsec. Gate calibration}

In the present experiment of emulating quantum information transmission, the native two-qubit CZ, $i$SWAP, and $\sqrt{i\rm SWAP}$ gates are used in order to reduce the circuit depth and execution time.

We implement nonadiabatic CZ gates using near-rectangular flux pulses with error-function-shaped rising and falling edges. The implementation protocol involves tuning qubit energy levels from their idle frequencies to operational frequencies, where the $|101\rangle$ and $|200\rangle$ (or $|002\rangle$) energy levels become resonant while activating the inter-qubit coupling. Here, the basis states are written in the order $|q_i,c_{ij},q_j\rangle$, where $c_{ij}$ denotes the tunable coupler connecting qubits $q_i$ and $q_j$.
Following sufficient phase accumulation during resonance, both the coupler and qubits return to their idle frequencies using the same smoothly shaped flux pulses.

Because the idle-frequency detuning between adjacent qubits is close to the qubit anharmonicity, the $|101\rangle$ and $|200\rangle$ (or $|002\rangle$) levels can be brought into resonance using a relatively small flux excursion. The nearly flat interaction plateau and smoothly shaped edges reduce sensitivity to waveform distortion and suppress nonadiabatic leakage. To suppress leakage and achieve the target conditional phase, we fine-tune the interaction frequencies, coupler flux amplitude, and interaction duration.
The calibrated CZ gate has a total duration of 61~ns, consisting of a 6~ns rising edge, a 45~ns interaction plateau, a 6~ns falling edge, and 4~ns of total buffer time.

The $i$SWAP and $\sqrt{i\rm SWAP}$ gates operate through coherent exchange between the $|001\rangle$ and $|100\rangle$ energy levels, using flux pulses with the same error-function-shaped edges. After bringing the qubit energy levels into resonance, the coupler is activated for a calibrated interaction duration that determines the
exchange angle. The calibrated $i$SWAP gate has a total duration of 46~ns, consisting of a 6~ns rising edge, a 30~ns interaction plateau, a 6~ns falling edge, and 4~ns of total buffer time.

An important consideration in implementing $i$SWAP and $\sqrt{i\rm SWAP}$ gates is the treatment of the single-qubit dynamical phases accumulated during the flux excursions. For the $i$SWAP gate, which produces a complete exchange of the two qubit excitations, these phases can be tracked and compensated virtually according to the phase-exchange rules described in Ref.~\cite{PhysRevLett.129.010502}. In contrast, the $\sqrt{i\rm SWAP}$ gate only partially exchanges the two excitation amplitudes, making direct virtual-frame tracking less convenient. To simplify the phase management protocol, we apply the required phase corrections before each $\sqrt{i\rm SWAP}$ gate using composite $R_z$ rotations constructed from physical $x$- and $y$-axis microwave pulses:
\begin{equation}
    R_z(\theta) = R_y\left(-\frac{\pi}{2}\right)R_x(\theta)R_y\left(\frac{\pi}{2}\right).
    \label{eq:composite-rz}
\end{equation}
Including these composite phase-correction rotations, the compiled $\sqrt{i\rm SWAP}$ gate has a total duration of 112~ns, of which 80~ns is associated with the single-qubit rotations and 32~ns with the entangling pulse sequence.

Gate performance for the CZ, $i$SWAP, and $\sqrt{i\rm SWAP}$ gates is evaluated using cross-entropy benchmarking (XEB), and the extracted gate fidelities and compiled gate durations are summarized in Table~\ref{tab:table2}.

\begin{table}[t]
	\centering
	\caption{Fidelity and gate length of two-qubit gates used in the present experiment.}
	\label{tab:table2}
	\setlength{\tabcolsep}{0.6cm}
	\begin{tabular}{c|ccccc|c}
		\hline \hline
		gate         & q$_r$-q$_1$   & q$_1$-q$_3$     & q$_1$-q$_2$   & q$_3$-q$_4$    & q$_2$-q$_4$    &
		gate length  \\
		\hline
		$i{\rm SWAP}$      & --         & $99.00\%$   & $99.16\%$  & $99.24\%$  & $99.13\%$   & $46\ \text{ns}$  \\
		$\sqrt{i{\rm SWAP}}$ & --         & $98.12\%$   & $98.64\%$  & $98.54\%$  & $97.9\%$    & $112\ \text{ns}$ \\
		CZ         & $99.53\%$  & $99.39\%$   & $99.57\%$  & --         & --          & $61\ \text{ns}$  \\
	\hline \hline
	\end{tabular}
\end{table}

\section{\theappsec. Derivation of Eq.~(2) in the main text}

	Equation~(2) in the main text gives the output of the quantum switch supermap $\mathcal{S}(\mathcal{E},\mathcal{F})$ on the input state $\ket{+}\bra{+}_{c}\otimes\rho$, here $\ket{\pm}_c\equiv(\ket{0}_c\pm i\ket{1}_c)/\sqrt{2}$. For Pauli channels, this output is exactly block diagonal in the $\{\ket{+}_c,\ket{-}_c\}$ basis, with the conditional target channels $\mathcal{C}_{\pm}$ and their probabilities $P_{\pm}$ appearing directly as the two diagonal blocks, expressed compactly in terms of Pauli commutators and anticommutators. Explicitly,
	\begin{align}
		\mathcal{S}(\mathcal{E}, \mathcal{F})(\ket{+}\bra{+}_{c} \otimes \rho) & = \ket{+}\bra{+}_{c}\otimes P_{+}\mathcal{C}_{+}(\rho) + \ket{-}\bra{-}_{c} \otimes P_{-}\, \mathcal{C}_{-}(\rho) ~, \label{eq-S:commun}\\
		\mathcal{C}_{\pm}(\rho) & = \frac{1}{4P_{\pm}}\sum_{i,j = 0}^{3} p_{i}q_{j}\,[\sigma_{i},\sigma_{j}]_{\pm}\,\rho\,[\sigma_{i},\sigma_{j}]_{\pm}^{\dagger}~, \nonumber
	\end{align}
	where $[A,B]_{\pm}=AB\pm BA$ and $P_{+}+P_{-}=1$.
	
	To derive Eq.~(\ref{eq-S:commun}) explicitly, we start from the Kraus representations of the two Pauli channels,
	\begin{equation}
		\mathcal{E}(\rho)=\sum_{i=0}^{3}p_i\,\sigma_i\rho\sigma_i,\qquad \mathcal{F}(\rho)=\sum_{j=0}^{3}q_j\,\sigma_j\rho\sigma_j,
	\end{equation}
	in terms of which the Kraus operators of the quantum switch read
	\begin{equation}
		W_{ij}=\sqrt{p_iq_j}\left(\sigma_i\sigma_j\otimes\ket{0}\bra{0}_{c}+\sigma_j\sigma_i\otimes\ket{1}\bra{1}_{c}\right).
		\label{eq-S:switch-kraus}
	\end{equation}
	The block-diagonal structure follows from a simple algebraic fact: any two Pauli operators either commute or anticommute, i.e., $\sigma_j\sigma_i=\eta_{ij}\sigma_i\sigma_j$ with $\eta_{ij}=+1$ ($-1$) for commuting (anticommuting) pairs. Acting on the input state, each Kraus operator therefore yields
	\begin{equation}
		W_{ij}\big(\ket{+}_{c}\otimes\ket{\psi}\big)=\sqrt{p_iq_j}\;\sigma_i\sigma_j\ket{\psi}\otimes\frac{\ket{0}_{c}+i\eta_{ij}\ket{1}_{c}}{\sqrt{2}},
		\label{eq-S:kraus-action}
	\end{equation}
	i.e., the control qubit is left in the definite state $\ket{+}_{c}$ ($\ket{-}_{c}$) whenever $\sigma_i$ and $\sigma_j$ commute (anticommute). Summing over all Kraus operators gives the manifestly block-diagonal output
	\begin{equation}
		\mathcal{S}(\mathcal{E},\mathcal{F})(\ket{+}\bra{+}_{c}\otimes\rho)=\sum_{i,j=0}^{3}p_iq_j\,\ket{\eta_{ij}}\bra{\eta_{ij}}_{c}\otimes\sigma_i\sigma_j\rho\sigma_j\sigma_i,
		\label{eq-S:quantum-switch-output}
	\end{equation}
	where $\ket{\eta_{ij}}_{c}=\ket{+}_{c}$ for $\eta_{ij}=+1$ and $\ket{\eta_{ij}}_{c}=\ket{-}_{c}$ for $\eta_{ij}=-1$.
	
	The two diagonal blocks define the unnormalized conditional channels $\tilde{\mathcal{C}}_{\pm}$. For the $\ket{+}$ block, only commuting pairs contribute, for which $[\sigma_i,\sigma_j]_{+}=2\sigma_i\sigma_j$, while anticommuting pairs can be included in the sum at no cost since $[\sigma_i,\sigma_j]_{+}=0$; the analogous statement holds for the $\ket{-}$ block. Hence
	\begin{align}
		\tilde{\mathcal{C}}_{+}(\rho)&=\frac{1}{4}\sum_{i,j=0}^{3}p_iq_j\,[\sigma_i,\sigma_j]_{+}\rho[\sigma_i,\sigma_j]_{+}^{\dagger},\label{eq-S:antichannel}\\
		\tilde{\mathcal{C}}_{-}(\rho)&=\frac{1}{4}\sum_{i,j=0}^{3}p_iq_j\,[\sigma_i,\sigma_j]_{-}\rho[\sigma_i,\sigma_j]_{-}^{\dagger}.\label{eq-S:commutator-channel}
	\end{align}
	
	The probabilities of the two branches are the traces of the unnormalized channels,
	\begin{equation}
		P_{\pm} = {\rm Tr}\bigl[\tilde{\mathcal{C}}_{\pm}(\rho)\bigr].
		\label{eq-S:trace-prob}
	\end{equation}
	For each anticommuting pair ($i,j\geq1$, $i\neq j$) one has $[\sigma_i,\sigma_j]_{-}=2\sigma_i\sigma_j$, so that the corresponding term contributes $p_iq_j\,{\rm Tr}[\sigma_i\sigma_j\rho\sigma_j\sigma_i]=p_iq_j$ to $P_{-}$, while commuting pairs contribute nothing. Evaluating the sum yields
	\begin{equation}
		P_{-} = (1 - p_{0})(1 - q_{0}) - \sum_{i=1}^{3} p_i q_i, \quad \text{and} \quad P_{+} = 1 - P_{-}.
		\label{eq-S:probabilities}
	\end{equation}
	The normalized channels are then obtained straightforwardly as $\mathcal{C}_{\pm}(\rho)=P_{\pm}^{-1}\,\tilde{\mathcal{C}}_{\pm}(\rho)$, which together with Eqs.~(\ref{eq-S:antichannel}), (\ref{eq-S:commutator-channel}), and (\ref{eq-S:probabilities}) completes the derivation of Eq.~(\ref{eq-S:commun}). 

\section{\theappsec. Quantum circuits for State Prep. and Tomo. in Fig. 2(a) of the main text}

\begin{figure}[t] 
	\centering 
	\includegraphics[width=0.7\textwidth]{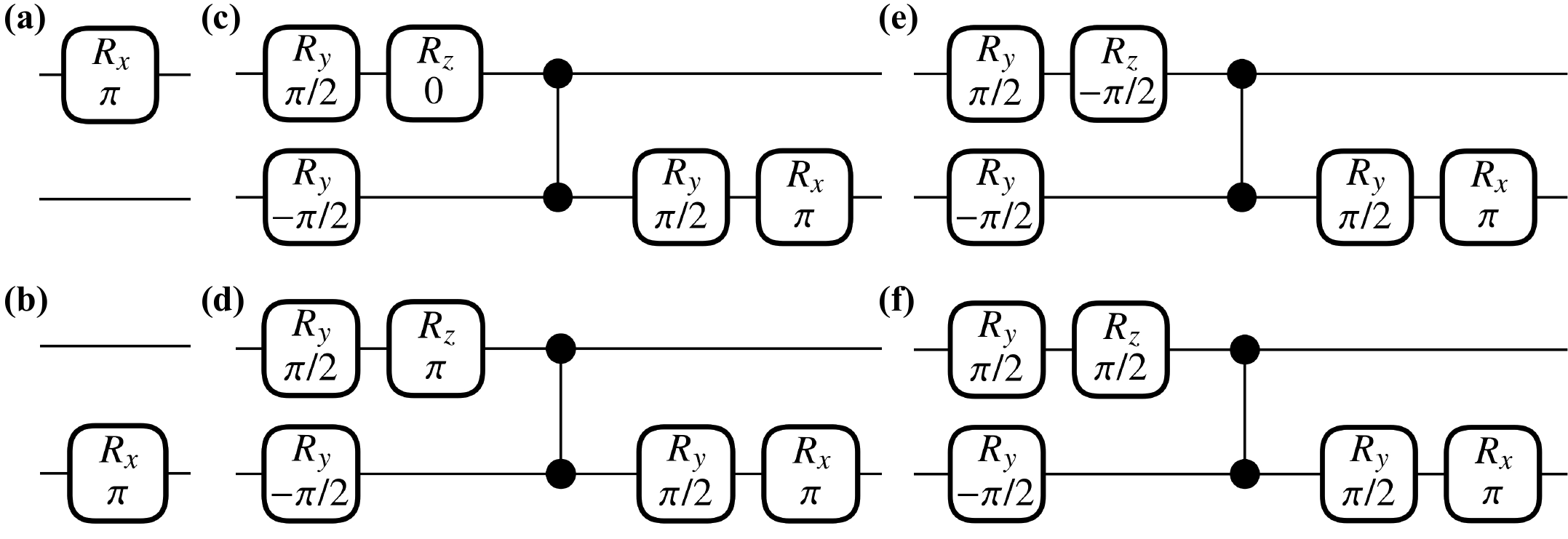}
	\caption{\centering Detailed circuits for the ''State Prep.'' block shown in Fig.~2(a) in the main text. The logical states $|0\rangle_L$, $|1\rangle_L$, $(|0\rangle_L+|1\rangle_L)/\sqrt{2}$, $(|0\rangle_L-|1\rangle_L)/\sqrt{2}$, $(|0\rangle_L+i|1\rangle_L)/\sqrt{2}$, and $(|0\rangle_L-i|1\rangle_L)/\sqrt{2}$ are prepared using the circuits in (a), (b), (c), (d), (e), and (f), respectively.}
	\label{fig:State Prep.} 
\end{figure}

Figure~\ref{fig:State Prep.} illustrates the detailed quantum circuits for the initial state preparation (State Prep.) block shown in Fig.~2(a) in the main text. The logical states $\ket{0}_L$, $\ket{1}_L$, $(\ket{0}_L+\ket{1}_L)/\sqrt{2}$, $(\ket{0}_L-\ket{1}_L)/\sqrt{2}$, $(\ket{0}_L+i\ket{1}_L)/\sqrt{2}$, and $(\ket{0}_L-i\ket{1}_L)/\sqrt{2}$ are prepared using the circuits in Figs.~\ref{fig:State Prep.}(a), (b), (c), (d), (e), and (f), respectively. These circuits employ single-qubit rotations ($R_x$, $R_y$, $R_z$) and controlled-$Z$ gates to map the physical states $\ket{10}$ and $\ket{01}$ onto the logical basis $\{\ket{0}_L, \ket{1}_L\}$, following the encoding scheme $\ket{0}_L \equiv \ket{01}$ and $\ket{1}_L \equiv \ket{10}$.

\begin{figure}[h] 
    \centering 
    \includegraphics[width=0.35\textwidth]{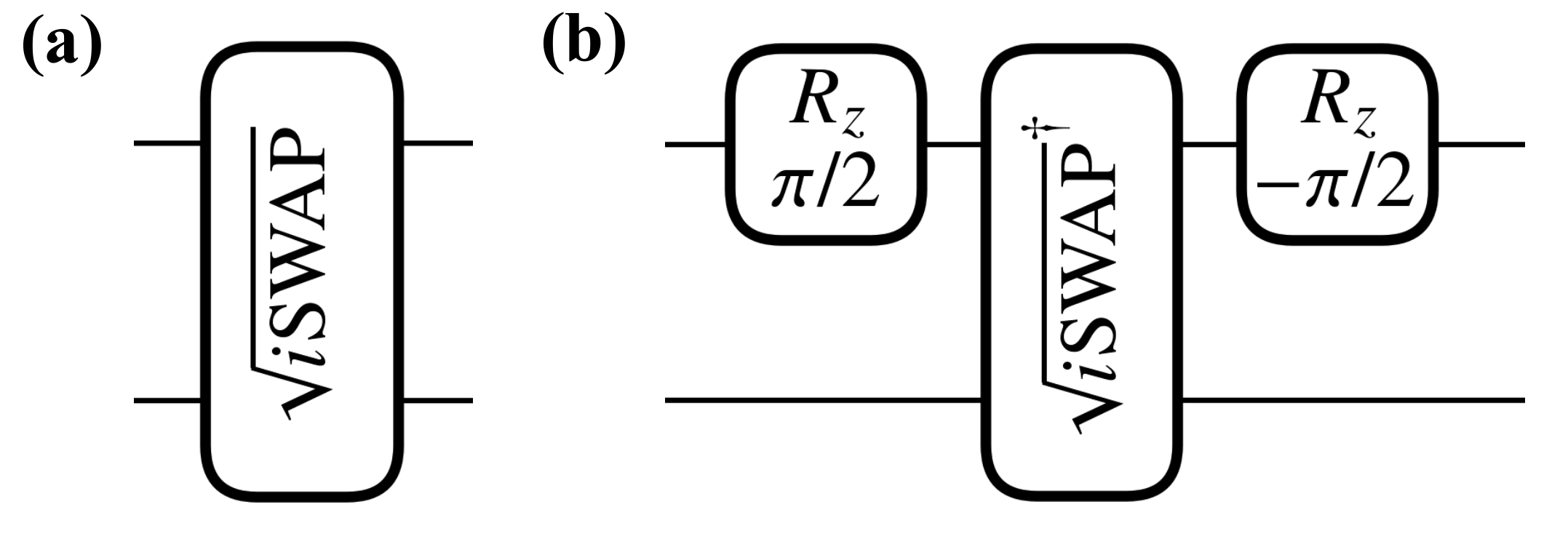}
    \caption{\centering Detailed circuits for the ''Tomo.'' block shown in Fig.~2(a) in the main text. (a) The logical $R_Y(-\pi/2)$ rotation for the measurement of logical $\langle$X$\rangle$. (b) The logical $R_X(\pi/2)$ rotation for the measurement of $\langle$Y$\rangle$.}
    \label{fig:tomo.} 
\end{figure}

The circuits for the Tomo. block in Fig. 2(a) in the main text are shown in Fig.~\ref{fig:tomo.}, which perform the necessary pre-rotations for measuring the logical Pauli observables. For $\langle Z\rangle$, no pre-rotation is required, since on the logical subspace $Z_L = Z\otimes I = -I\otimes Z$, a direct $Z$-basis measurement of one physical qubit (with the appropriate sign) yields $\langle Z_L\rangle$. Figure~\ref{fig:tomo.}(a) shows the circuit for measuring $\langle X \rangle$, which is a logical $R_Y(-\pi/2)$ rotation applied before the final projective measurement. {\it Here and below, $R_Y$ with the subscript in upper case denotes a rotation of the logical qubit, in contrast to $R_y$ for the physical qubit}. Figure~\ref{fig:tomo.}(b) presents the circuit for measuring $\langle Y \rangle$, where an $R_X(\pi/2)$ rotation is applied prior to readout. These pre-measurement rotations map the logical Pauli operators onto the computational basis, enabling direct extraction of the desired observables. Together, the State Prep. and Tomo. circuits constitute the complete sequence for quantum state initialization and characterization in the present experiment.

\section{\theappsec. Post selection for the reconstructed matrix}
\label{sec:post-selection-ptm}

As shown in Fig.~2(a) and Fig.~3(b) in the main text, we consider the ICO channel $\mathcal{S}(\mathcal{E},\mathcal{F})$, where $\mathcal{E}(\rho)=\sum_i p_i\sigma_i\rho\sigma_i$ and $\mathcal{F}(\rho)=\sum_j q_j\sigma_j\rho\sigma_j$ are two noisy Pauli channels. The first two $\sqrt{\mathrm{iSWAP}}$ gates act as a $50:50$ beam splitter, splitting the state in $Q_1$ into a coherent superposition across $Q_1$ and $Q_2$. The two branches pass through the two channels in opposite orders, with the Pauli operators $\sigma_i$ and $\sigma_j$ applied in reversed sequences. During this stage, the control qubit of the quantum switch is encoded in which the logical qubit carries the excitation: the control qubit is in the $\ket{0}_c$ ($\ket{1}_c$) state when the excitation resides in $Q_1$ ($Q_2$). The last two $\sqrt{\mathrm{iSWAP}}$ gates recombine the two branches, mapping $\ket{-}_c$ and $\ket{+}_c$ back to $Q_1$ and $Q_2$ with $\ket{\pm}_c\equiv(\ket{0}_c\pm i\ket{1}_c)/\sqrt{2}$, respectively.

For the reconstructed channel matrix $\gamma_{ij}$, we note that the logical qubit subspace is spanned by $\ket{0}_L \equiv \ket{01}$ and $\ket{1}_L \equiv \ket{10}$, while the states $\ket{\rm vac} \equiv \ket{00}$ and $\ket{11}$ are outside this subspace. Ideally, the emulation of quantum communication will lie within the logical qubit subspace. Experimentally, however, the qubit decoherence, gate infidelity, and possible state leakage will lead the process out of the subspace, and post selection is required for the final data analysis. Let \(P_{01}\) and \(P_{10}\) denote the probabilities of measuring the two physical qubits encoding the logical qubit in \(|0\rangle_L\) and \(|1\rangle_L\), respectively. In the post-selected data analysis, only these two classes of events are kept. The post-selection probability is
\begin{equation}
P=P_{01}+P_{10}.
\label{eq:ptm-Pps}
\end{equation}
On the logical qubit subspace, the corresponding probabilities are
\begin{equation}
p_{01}=\frac{P_{01}}{P},\qquad
p_{10}=\frac{P_{10}}{P} .
\label{eq:ptm-p}
\end{equation}
Logical expectation values are then computed within this subspace. For example, the logical \(Z\) expectation value is
\begin{equation}
\langle Z\rangle
= p_{01}-p_{10}
= \frac{P_{01}-P_{10}}{P_{01}+P_{10}}~.
\label{eq:ptm-Z}
\end{equation}
Analogous post-selected expectation values are obtained for the $\langle X\rangle$ and $\langle Y\rangle$ bases and are used to construct \(\gamma_{ij}\). When readout calibration is available, the measured four-bit probabilities are first corrected by the readout matrix, and the above logical-subspace post selection is applied afterwards.

According to Eq.~(2) in the main text, the ICO channel $\mathcal{S}(\mathcal{E},\mathcal{F})$ acts on the target state $\rho$ through two branches, and conditioned on the control qubit being in $\ket{\pm}_c$, the target undergoes the effective channel $\mathcal{C}_{\pm}$ with probability $P_{\pm}$. For the three quantum switch channels shown in Fig.~2 of the main text, the post-selection probabilities \(P_{\pm}\) defined in Eq.~(2) of the main text are summarized in Table~\ref{tab:postsel}. The theoretical values are calculated from Eq.~(2) of the main text using the corresponding channel parameters, while the experimental values are obtained as $P_\pm^{\mathrm{exp}} / (P_+^{\mathrm{exp}} + P_-^{\mathrm{exp}})$ from the measurement (see also below). The measured \(P_{\pm}\) agree well with the theoretical predictions, with absolute deviations below \(0.02\). These probabilities are used to normalize the channels \(\mathcal{C}_{\pm}(\rho)\).

\section{\theappsec. Post selection for quantum capacity}
\label{sec:post-selection-quantum-capacity}

\begin{table}[t]
	\centering
	\caption{\centering Comparison of experimental and theoretical post-selection probabilities \(P_{+}\) and \(P_{-}\) for the three quantum switch channels in Fig.~2 of the main text: \(\mathcal{S}(\mathcal{P}_1,\mathcal{P}'_1)\) with \(s=t=0.5\), \(\mathcal{S}(\mathcal{P}_2,\mathcal{P}_2)\) with \(p_1=p_2=0.5\), and  \(\mathcal{S}(\mathcal{P}_3,\mathcal{P}_3)\) with \(p_1=p_2=p_3=1/3\). Theoretical values are calculated from Eq.~(2) of the main text. Experimental values are extracted from the control-qubit measurements in the \(|\pm\rangle_c\) basis and normalized such that \(P_{+}+P_{-}=1\). The absolute deviations between experiment and theory are below \(0.02\).}
	\label{tab:postsel}
	\setlength{\tabcolsep}{0.4cm}
	\begin{tabular}{c|cc|cc|cc}
		\hline \hline
		& $\mathcal{S}(\mathcal{P}_1,\mathcal{P}'_1)_{\rm exp}$ &  $\mathcal{S}(\mathcal{P}_1,\mathcal{P}'_1)_{\rm theory}$   & $\mathcal{S}(\mathcal{P}_2,\mathcal{P}_2)_{\rm exp}$   &  $\mathcal{S}(\mathcal{P}_2,\mathcal{P}_2)_{\rm theory}$  &  $\mathcal{S}(\mathcal{P}_3,\mathcal{P}_3)_{\rm exp}$  &  $\mathcal{S}(\mathcal{P}_3,\mathcal{P}_3)_{\rm theory}$ \\
		\hline
		$P_{+}$      & 0.744         &  0.750 & 0.509 & 0.500  &  0.353  & 0.333 \\  
		$P_{-}$    & 0.256         & 0.250  & 0.491 &  0.500 &  0.647  & 0.667\\
		\hline \hline
	\end{tabular}
\end{table}

For the coherent information $I_Q$ of the ICO channel $\mathcal{S}(\mathcal{E},\mathcal{F})$, the circuit Fig.~3(b) in the main text contains a reference qubit $q_r$ together with the  two logical qubits $Q_1$ and $Q_2$. Initially, $q_r$ and $Q_1$ carry the purification $\ket{\Phi_\rho}$ of the target mixed state $\rho$, while $Q_2$ is in the vacuum state. Due to the block-diagonal structure in the basis $\{\ket{\pm}_c\}$, the coherent information decomposes as $I_Q=P_+I_Q(\mathcal{C}_+)+P_-I_Q(\mathcal{C}_-)$, with
\begin{equation}
	I_Q(\mathcal{C}_\pm) = S_{vN}(\rho_\pm)-S_{vN}(\omega_\pm). \label{eq:I_QCpm}
\end{equation}
Here we denote the channel output and the joint reference-output states as $\rho_\pm\equiv\mathcal{C}_\pm(\rho)$ and $\omega_\pm\equiv\mathcal{I}_r\otimes\mathcal{C}_\pm\left(\ket{\Phi_\rho}\bra{\Phi_\rho}\right)$, respectively. Since the last beam splitter maps $\ket{-}_c$ ($\ket{+}_c$) to $Q_1$ ($Q_2$), the output states $\omega_\pm$ reside in the Hilbert spaces of $q_rQ_1$ and $q_rQ_2$, respectively. For each Kraus pair $(\sigma_i,\sigma_j)$, we post-select on the cases where the excitation is detected in $Q_1$ (for $\mathcal{C}_-$) or in $Q_2$ (for $\mathcal{C}_+$), corresponding to the control qubit being projected onto $\ket{-}_c$ or $\ket{+}_c$, respectively. After this post selection, we reconstruct the joint density matrices $\rho^{ij}_{q_rQ_1}$ and $\rho^{ij}_{q_rQ_2}$ from the retained counts by maximum-likelihood quantum state tomography~\cite{PhysRevLett.108.070502}. We project the resulting density matrices onto the single-excitation subspaces of $Q_1$ and $Q_2$ using the projectors $\Pi_{Q_1}$ and $\Pi_{Q_2}$, respectively, and average the projected states with weights $p_iq_j$ over all Kraus pairs:
\begin{equation}
	\tilde{\omega}_- = \sum_{i,j}p_iq_j\Pi_{Q_1}\rho^{ij}_{q_rQ_1}\Pi_{Q_1},\quad\tilde{\omega}_+ = \sum_{i,j}p_iq_j\Pi_{Q_2}\rho^{ij}_{q_rQ_2}\Pi_{Q_2}.
\end{equation}
Anticommuting pairs give vanishing contributions to $\tilde{\omega}_+$ because their excitations are routed to $Q_1$ and discarded by the $Q_2$ post selection. Likewise, commuting pairs contribute zero to $\tilde{\omega}_-$. The branch probabilities are defined operationally as $P_\pm = \mathrm{Tr}[\tilde{\omega}_\pm]$. In the ideal noiseless case this coincides with the theoretical expression $P_\pm = \sum_{(i,j)\in\pm} p_i q_j$, ensuring $P_+ + P_- = 1$ exactly. In the experiment, however, decoherence, gate imperfections, and readout errors cause the measured values to deviate from this ideal behavior, yielding $P_+^{\mathrm{exp}} + P_-^{\mathrm{exp}} \neq 1$. The deviation from unity provides a quantitative measure of the total error magnitude. To maintain well-defined probabilities, we renormalize the experimental branch probabilities so that $P_\pm$ is given by $P_\pm^{\mathrm{exp}} / (P_+^{\mathrm{exp}} + P_-^{\mathrm{exp}})$, and the normalized conditional states are then $\omega_\pm = P_\pm^{-1} \tilde{\omega}_\pm$ (see the first column in Table~\ref{tab:postsel} for the experimentally determined, renormalized branch probabilities $P_\pm$ alongside their ideal theoretical values $P_\pm=\sum_{(i,j)\in\pm}p_iq_j$ with good agreement). Finally, tracing out $q_r$ yields $\rho_\pm=\mathrm{Tr}_{q_r}[\omega_\pm]$. From $\rho_\pm$ and $\omega_\pm$ we evaluate the conditional coherent information $I_Q(\mathcal{C}_\pm)=S_{vN}(\rho_\pm)-S_{vN}(\omega_\pm)$, and the total coherent information follows as $I_Q=P_+I_Q(\mathcal{C}_+)+P_-I_Q(\mathcal{C}_-)$.

In the ideal noiseless limit, the post selection on the vacuum state of the idle logical qubit is not strictly necessary, because the final beam splitter already routes commuting and anticommuting Kraus pairs to $Q_2$ and $Q_1$, respectively, The single-excitation projector on the active pair therefore automatically discards the opposite branch. Its primary role in the experiment is error mitigation. By first measuring the idle pair and retaining only shots in the $\ket{00}$ state, we reject events in which the total excitation number has leaked outside the single-excitation manifold --- for instance, due to double excitations or qubit relaxation. The subsequent application of $\Pi_{Q_1}$ or $\Pi_{Q_2}$ then enforces that the reconstructed conditional state is strictly supported on the single-excitation subspace of the active logical qubit.

\section{\theappsec. Channels and information loss}

Quantum channels model the degradation of quantum information due to interaction with the environment. In this work, we focus on four typical noisy channels: the bit‑flip channel $\mathcal{P}_1(\rho) = (1-s)\rho + s\sigma_1\rho\sigma_1$, the phase‑flip channel $\mathcal{P}'_1(\rho) = (1-t)\rho + t\sigma_3\rho\sigma_3$, and the Pauli channels $\mathcal{P}_2(\rho) = \sum_{i=1}^2 p_i\sigma_i\rho\sigma_i$ and $\mathcal{P}_3(\rho) = \sum_{i=1}^3 p_i\sigma_i\rho\sigma_i$. In the experiment, the channels ${\mathcal P}_1{\mathcal P}'_1$ and ${\mathcal P}'_1{\mathcal P}_1$ are decomposed into four components ($II$, $IZ$, $XI$, $XZ$) and ($II$, $IX$, $ZI$, $ZX$), respectively. Similarly, channels ${\mathcal P}_2{\mathcal P}_2$ and ${\mathcal P}_3{\mathcal P}_3$ have the components ($XX$, $XY$, $YX$, $YY$) and ($XX$, $XY$, $XZ$, $YX$, $YY$, $YZ$, $ZX$, $ZY$, $ZZ$), respectively.

\begin{figure}[tp] 
	\centering 
	\includegraphics[width=0.6\textwidth]{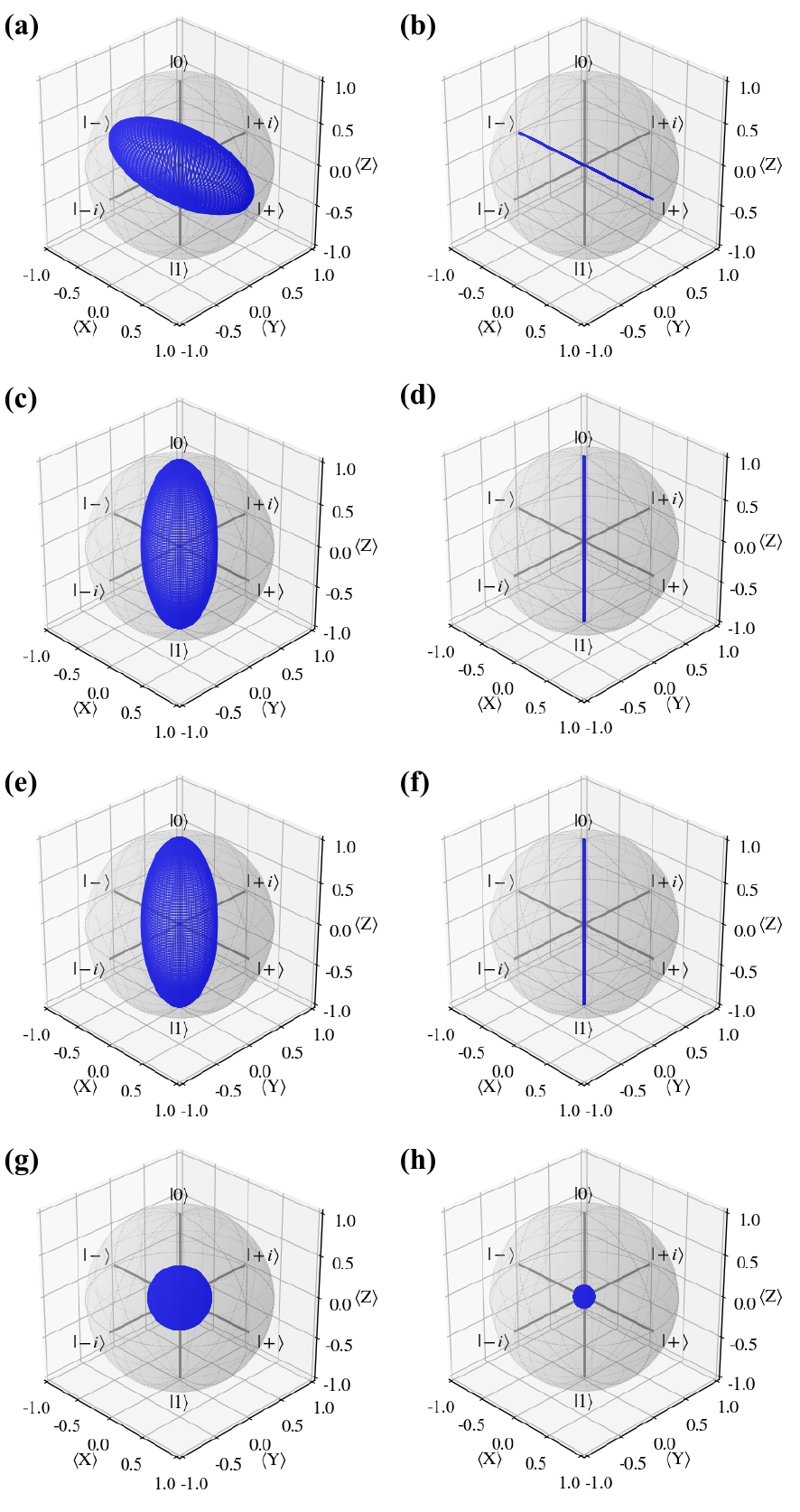}
	\caption{\centering Noise effects of various quantum channels seen on the Bloch sphere. Bit‑flip channel $\mathcal{P}_1(\rho) = (1-s)\rho + s\sigma_1\rho\sigma_1$ for (a) $s=0.3$ and (b) $s=0.5$. Phase‑flip channel $\mathcal{P}'_1(\rho) = (1-t)\rho + t\sigma_3\rho\sigma_3$ for (c) $t=0.3$ and (d) $t=0.5$. Pauli channel $\mathcal{P}_2(\rho) = \sum_{i=1}^2 p_i\sigma_i\rho\sigma_i$ with parameters (e) $p_1=0.3, p_2=0.7$ and (f) $p_1=p_2=0.5$.  (g) Pauli channel $\mathcal{P}_3(\rho) = \sum_{i=1}^3 p_i\sigma_i\rho\sigma_i$ with $p_1=p_2=p_3=1/3$. (h) Two consecutive applications of $\mathcal{P}_3\mathcal{P}_3$ with $p_1=p_2=p_3=1/3$.} 
	\label{fig:channel} 
\end{figure}

As illustrated in Figs.~\ref{fig:channel}(a) through (h), these channels contract the Bloch sphere along specific axes. For instance, the bit‑flip channel $\mathcal{P}_1$ collapses the sphere gradually to the $x$‑axis from $s=0.3$ in (a) to $s=0.5$ in (b), completely destroying quantum coherence (entanglement breaking). Similarly, the phase‑flip channel $\mathcal{P}'_1$ compresses the state onto the $z$‑axis with increasing $t$ toward 0.5. The Pauli channels $\mathcal{P}_2$ and $\mathcal{P}_3$ induce simultaneous contraction along two or three axes, leading to severe information loss as shown in Figs.~\ref{fig:channel}(e, f) and (g, h), respectively. Particularly, if two of these channels are concatenated in a fixed order, the quantum information will be greatly or entirely lost during transmission.

\section{\theappsec. Quantum circuits for quantum Hadamard transformation}

\begin{figure}[t] 
	\centering 
	\includegraphics[width=0.6\textwidth]{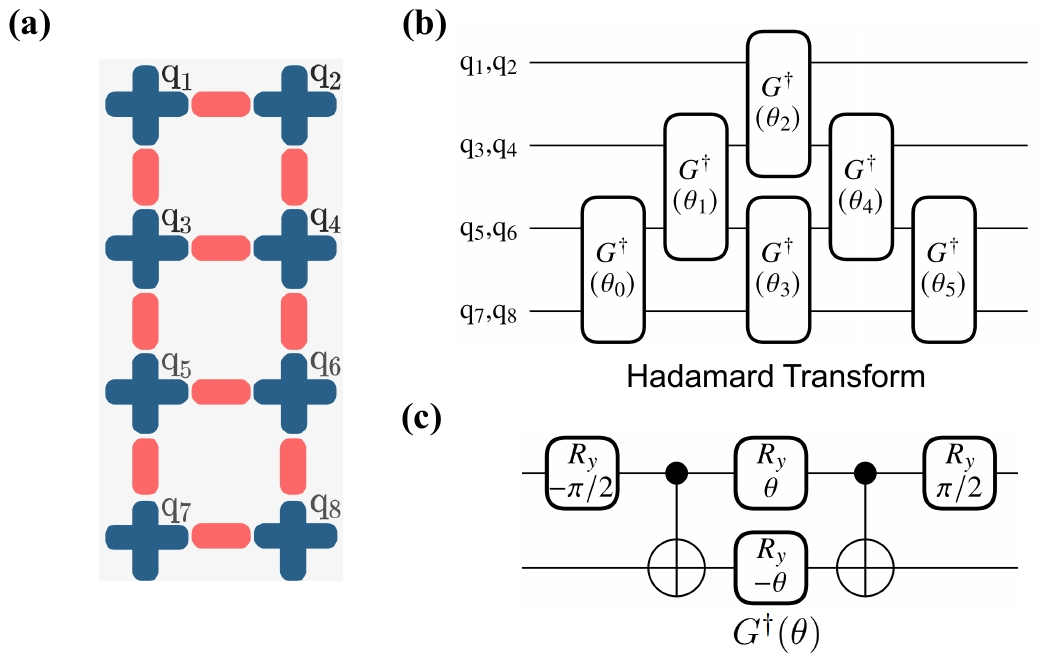}
	\caption{\centering Compiling circuit for the quantum Hadamard transformation with logical qubits. (a) Schematic of the eight-qubit ladder-layout processor. (b) Implementation of the four-dimensional Hadamard transform $H_4$ in Eq.~\eqref{eq:H4} in the single-excitation subspace via a sequence of Givens rotations, each labeled by its angle $\theta_i$, with $i=0,1,...N_G-1$. (c) Elementary two-qubit realization of a single Givens rotation $G^{\dagger}(\theta)$ using only $R_y$ rotational and CNOT gates.} 
	\label{fig:gate order} 
\end{figure}

One of the ingredients for implementing the HPP algorithm is a systematic way to construct the quantum Hadamard transformation (QHT). In the photonic experiment, the 4-dimensional QHT is implemented by a 4-core fiber beam splitter. In superconducting qubit systems, we implement arbitrary \(N\)-dimensional QHTs in the single-excitation subspace spanned by the basis states
\begin{equation}
\left\{ \ket{1}_{L,j} \coloneqq \ket{0\ldots 1_j\ldots 0}_L \right\}_{j=0}^{N-1}
\end{equation}
through a linear-depth Givens-rotation circuit. This circuit architecture has previously been proposed to prepare arbitrary Slater determinants~\cite{knill2001fermionic, PhysRevApplied.9.044036}. The key idea is to treat the Hadamard matrix as a linear transformation between two sets of fermionic modes, i.e.,
\begin{equation}
b_{x}^{\dagger} = \frac{1}{\sqrt{N}} \sum_{y} H_{x,y} a_{y}^{\dagger},
\end{equation}
where \(a_{y}^{\dagger}\) and \(b_{x}^{\dagger}\) are the fermionic creation operators before and after the transformation. The Givens rotation can be parameterized by a single angle \(\theta\), i.e.,
\begin{equation}
G(\theta) = \begin{pmatrix}
\cos\theta & -\sin\theta \\
\sin\theta & \cos\theta
\end{pmatrix}.
\label{eq:A2}
\end{equation}
When a Givens rotation is applied to a pair of fermionic modes denoted as \((a_{p},a_{q})\), the corresponding unitary operation is
\begin{equation}
R_{G}(\theta) = e^{\theta (a_{p}^{\dagger}a_{q} - a_{q}^{\dagger}a_{p})}.
\end{equation}
The scaled Hadamard matrix \(H_N = M_N / \sqrt{N}\) can be decomposed into a sequence of Givens rotations and a diagonal matrix \(\Gamma\), i.e.
\begin{equation}
H_N = \Gamma\, G^{\dagger}(\theta_{N_G-1}) \cdots G^{\dagger}(\theta_{0}),
\end{equation}
with \(\Gamma_{j,k} = \pm 1\,\delta_{j,k}\) and the number of Givens rotations \(N_G = N(N - 1)/2\). Note that in the above equation, each Givens rotation is embedded into a \(N\times N\) identity matrix. By introducing the Jordan-Wigner transformation and requiring that each Givens rotation acts on adjacent fermionic modes, the QHT can be decomposed into a quantum circuit with linear connectivity and nearest-neighbor two-qubit gates. Before the Givens rotations, a Pauli-\(Z\) gate is applied to the \(j\)-th qubit if \(\Gamma_{j,j} = -1\) to maintain the phase relations.

\begin{figure}[tp] 
	\centering 
	\includegraphics[width=0.85\textwidth]{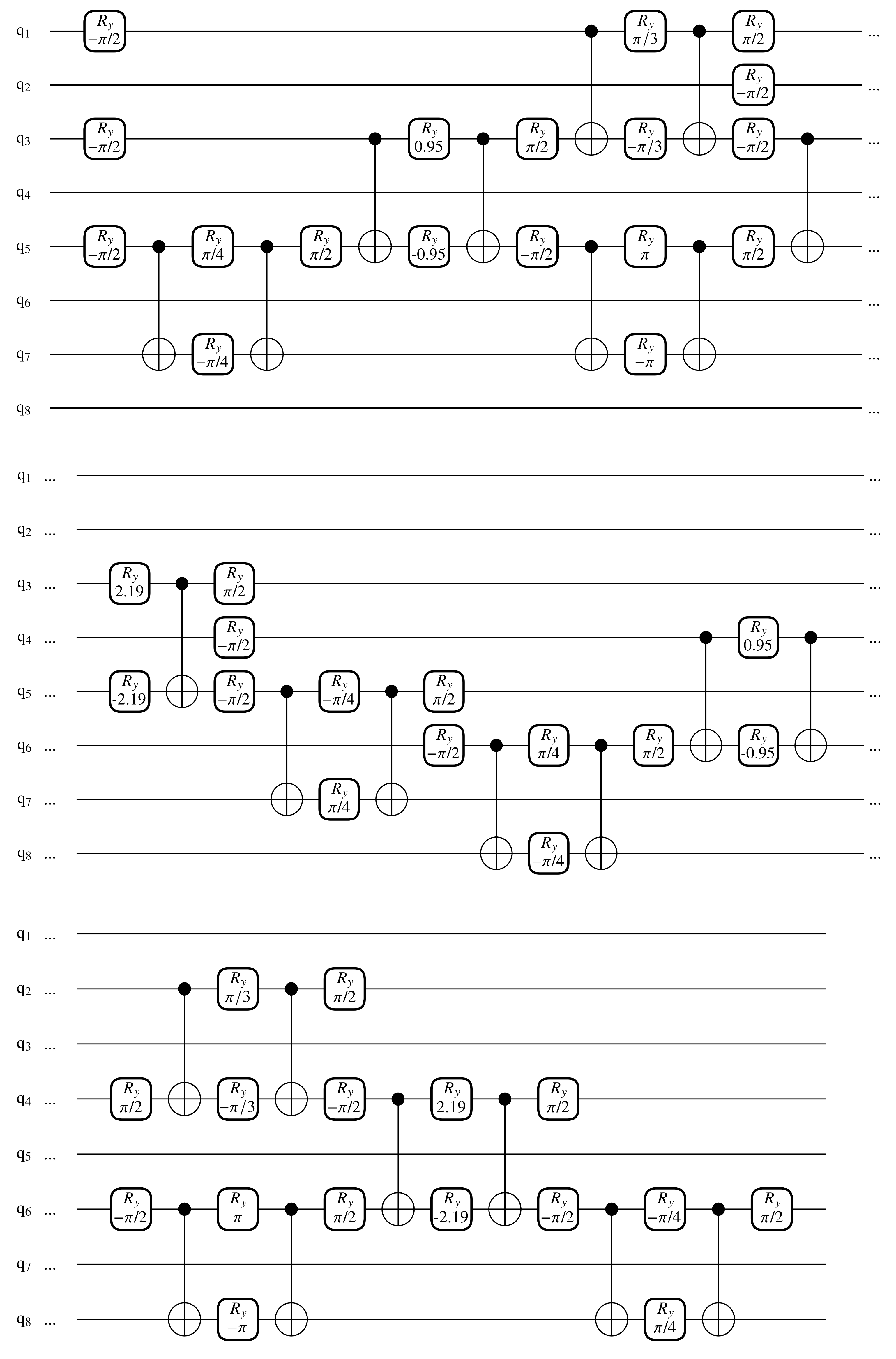}
	\caption{\centering Complete compiled circuit for the four logical-qubit Hadamard transform.} 
	\label{fig:all_hpp} 
\end{figure}

\begin{figure}[t] 
	\centering 
	\includegraphics[width=0.45\textwidth]{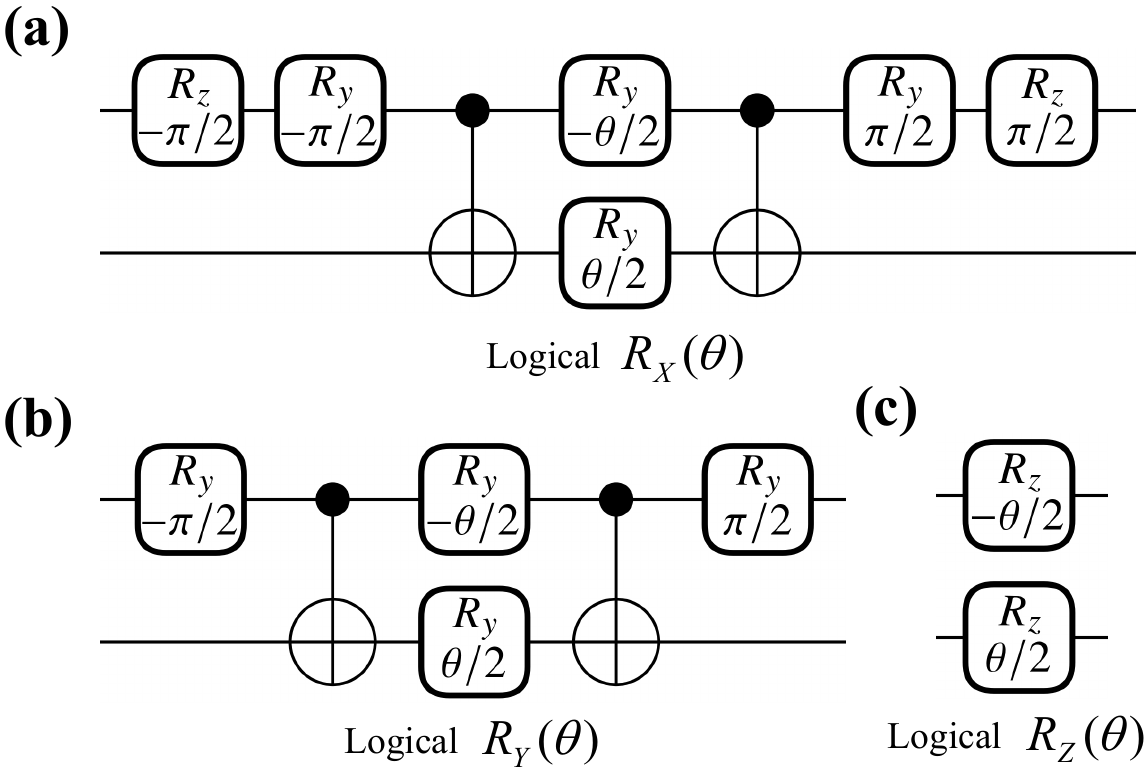}
	\caption{\centering Elementary circuits for logical qubits. (a) $R_{X}(\theta)$. (b) $R_{Y}(\theta)$. (c) $R_{Z}(\theta)$.} 
	\label{fig:Logical_gate} 
\end{figure}

As a concrete illustration of the above construction, we now present the compiled decomposition used in the HPP simulation. Figures~\ref{fig:gate order} and \ref{fig:all_hpp} summarize the compiled decomposition of the quantum Hadamard transformation for $N=4$. The ladder-layout processor Fig.~\ref{fig:gate order}(a) consists of eight physical qubits arranged as four logical qubits.  For each logical qubit, the vacuum state and the logical computational basis states are
\begin{equation}
	\ket{\mathrm{vac}}\coloneqq\ket{00},\quad \ket{0}_L\coloneqq\ket{01},\quad \ket{1}_L\coloneqq\ket{10},
\end{equation}
and the logical Pauli operators $\{I, X, Y, Z\}$ are defined in the logical computational space, i.e.,
\begin{equation}
	I = \ket{01}\bra{01}+\ket{10}\bra{10},~X = \ket{01}\bra{10}+\ket{10}\bra{01},~Y = -i\ket{01}\bra{10}+i\ket{10}\bra{01},~Z = \ket{01}\bra{01}-\ket{10}\bra{10}.
\end{equation}
Arbitrary single-qubit logical operations can be decomposed into logical Pauli rotations $R_X(\theta)$, $R_Y(\theta)$ and $R_Z(\theta)$, which can be further compiled with single-qubit rotations and CNOT gates, as shown in Fig.~\ref{fig:Logical_gate}. The unitary operations of $R_X(\theta)$, $R_Y(\theta)$ and $R_Z(\theta)$ in the computational basis of the two physical qubits are
\begin{equation}
	\begin{aligned}
		R_X(\theta) & = \ket{00}\bra{00}+\ket{11}\bra{11}+\cos\frac{\theta}{2}\left(\ket{01}\bra{01}+\ket{10}\bra{10}\right)-i\sin\frac{\theta}{2}\left(\ket{01}\bra{10}+\ket{10}\bra{01}\right),\\
		R_Y(\theta) & = \ket{00}\bra{00}+\ket{11}\bra{11}+\cos\frac{\theta}{2}\left(\ket{01}\bra{01}+\ket{10}\bra{10}\right)-\sin\frac{\theta}{2}\left(\ket{01}\bra{10}-\ket{10}\bra{01}\right),\\
		R_Z(\theta) & = \ket{00}\bra{00}+\ket{11}\bra{11}+\cos\frac{\theta}{2}\left(\ket{01}\bra{01}+\ket{10}\bra{10}\right)-i\sin\frac{\theta}{2}\left(\ket{01}\bra{01}-\ket{10}\bra{10}\right).
	\end{aligned}
\end{equation}
It is clear from the above equation that $R_\alpha(\theta)$ ($\alpha=X,Y,Z$) act as logical Pauli rotations in the logical computational space while leaving the vacuum state unchanged.

The complete quantum circuits for the HPP simulation with indefinite causal orders are shown below. For the explicit construction for $N=K=4$ considered in the main text, we use the Hadamard transform in Ref.~\cite{PRXQuantum.2.010320} for convenience, given by

\begin{equation}
H_4 = \frac{1}{2}
\begin{pmatrix}
1 & 1 & 1 & 1 \\
1 & 1 & -1 & -1 \\
1 & -1 & -1 & 1 \\
1 & -1 & 1 & -1
\end{pmatrix}.
\label{eq:H4}
\end{equation}

Applying the Givens decomposition procedure described above yields the sequence of rotations shown in Fig.~\ref{fig:gate order}(b). For the particular case of $H_4$, the reduction yields $\Gamma = {\mathbb I}_4$ and six Givens rotations on adjacent modes. Denoting the four modes by $j=0,1,2,3$, the ordered sequence of rotations is on mode pairs $(2,3)$, $(1,2)$, $(0,1)$, $(2,3)$, $(1,2)$, and $(2,3)$, with corresponding angles

\begin{equation}
\theta_0 = \frac{\pi}{4},\quad
\theta_1 = \arctan(\sqrt{2}) \approx 0.95,\quad
\theta_2 = \frac{\pi}{3},\quad
\theta_3 = \pi,\quad
\theta_4 = \pi-\arctan(\sqrt{2}) \approx 2.19,\quad
\theta_5 = -\frac{\pi}{4}.
\end{equation}
These angles exactly realize the matrix $H_4$ in Eq.~(\ref{eq:H4}), which exhibit a symmetry under the Givens reduction: $\theta_0$ and $\theta_5$ form an opposite-sign pair ($\pm\pi/4$), while $\theta_1$ and $\theta_4$ are complementary ($\theta_1+\theta_4=\pi$). Both relations are a direct consequence of the symmetry of the orthogonal Givens reduction applied to this particular Hadamard matrix. Each rotation $G^{\dagger}(\theta_i)$, with $i=0,1,...N_G-1$ is implemented using the $R_y$ and CNOT decomposition in Fig.~\ref{fig:gate order}(c).

The complete circuit is shown in Fig.~4(a) in the main text. It implements a quantum switch that solves the Hadamard promise problem, where the Hadamard transform is given by the matrix $H_4$ in Eq.~(\ref{eq:H4}). The Hadamard operation on the four logical modes decomposes into two identical copies acting on the odd-indexed physical qubits $(q_1,q_3,q_5,q_7)$ and the even-indexed qubits $(q_2,q_4,q_6,q_8)$, as illustrated in Fig.~\ref{fig:gate order}(b). In the ladder-layout geometry shown in Fig.~\ref{fig:gate order}(a), both transforms can be executed simultaneously.

\section{\theappsec. Hadamard promise problems with unitaries in random basis}

In Table~I of the main text, we choose the gate sets from pure Pauli operators $U_j \in \{I, X, Y, Z\}$, picking one representative set per promise class from the 136 pure-Pauli sets that satisfy the Hadamard promise for $N=K=4$. The elementary logical Pauli rotations in Fig.~\ref{fig:Logical_gate}, namely $R_X(\theta)$, $R_Y(\theta)$, and $R_Z(\theta)$, serve as the building blocks for all the compiled circuits. A bare Pauli gate is implemented directly by a single logical $\pi$-rotation: each of $U_i = X, Y, Z$ is realized by the corresponding rotation $R_X(\pi)$, $R_Y(\pi)$, or $R_Z(\pi)$, while $U_i = I$ corresponds to the identity. Together with the quantum Hadamard transformation discussed in the previous section and its inverse, these gate realizations provide a complete decomposition of the HPP simulation in Fig.~4(a) in the main text into elementary logical operations.

To further explore the advantage of ICO for solving the Hadamard promise problem, Ref.~\cite{PRXQuantum.2.010320} proposed to randomize the basis in which each gate set is delivered by conjugating all four unitaries with a Haar-random single-qubit rotation $W$, i.e., $\mathsf{U}^{(W)} = \{W U_i W^{\dagger}\}_{i=1}^{K}$. The Hadamard promise is invariant under this collective conjugation, so the rotated set $\mathsf{U}^{(W)}$ satisfies the promise with the same column index $y$, and any strategy predicated on recognizing the specific operator form of the gates is rendered ineffective because measurements can no longer be aligned with the eigenbases of the delivered unitaries. Each bare gate $U_i$ is an involutory Pauli operator, so its conjugate $W U_i W^{\dagger}$ is a $\pi$-rotation about a certain Pauli axis of $U_i$. This $\pi$-rotation about a generic axis is compiled through the Euler decomposition in the $YZY$ convention:
	\begin{equation}
		W U_i W^{\dagger} = R_Y(\gamma_i)\, R_Z(\beta_i)\, R_Y(\alpha_i),
	\end{equation}
	up to a global phase, where the Euler angles $\alpha_i, \gamma_i \in [0, 2\pi)$ and $\beta_i \in [0, \pi]$ are determined by the axis and hence by both the random rotation $W$ and the gate $U_i$. The Euler angles can be programmed directly at the pulse level by composing the elementary logical rotations in Fig.~\ref{fig:Logical_gate}, so that a conjugated gate is compiled from up to three logical rotations instead of the single rotation required for a bare Pauli gate. This continuous tunability makes superconducting platforms particularly well suited for studying quantum control of gate orders in the Hadamard promise problem with randomized basis.

\begin{figure}[t] 
	\centering 
	\includegraphics[width=0.55\textwidth]{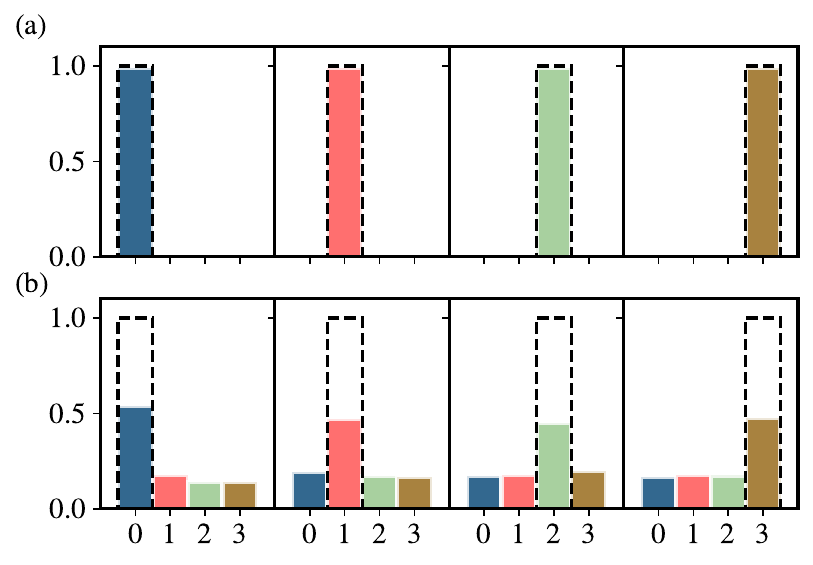}
	\caption{\centering Averaged output $y$-distribution of the random-rotation Hadamard promise problem for the four promise classes $y=0,1,2,3$ (columns), obtained by numerical simulation for the pure-Pauli gate sets in Table~I of the main text conjugated by Haar-random single-qubit rotations and post-selected on the single-excitation subspace. Within each panel, the bars give the probability of each output $y$ (horizontal axis), averaged over 20 random rotations, with the dashed outline marking the ideal distribution. (a) $T_1=100\,\mu\mathrm{s}$, $T_2=75\,\mu\mathrm{s}$. (b) $T_1=20\,\mu\mathrm{s}$, $T_2=2\,\mu\mathrm{s}$.}
	\label{fig:decoherence} 
\end{figure}

The numerical simulation in Fig.~4(b) of the main text employs a thermal-relaxation noise model, characterized by the longitudinal relaxation time $T_1$ and transverse relaxation time $T_2$ applied over a gate duration of $30$~ns for both single-qubit $R_y$ and $R_z$ gates and two-qubit CNOT gate. Two representative coherence settings are considered: (i) $(T_1,T_2)=(100,75)\,\mu\mathrm{s}$, achievable by state-of-the-art superconducting processors in Ref.~\cite{Abanin2025}; (ii) $(T_1,T_2)=(20,2)\,\mu\mathrm{s}$, comparable to the coherence times of the device used in this work, see Table~\ref{tab:table1}. Setting (i) is used in the noisy simulation of Fig.~4(b) in the main text, which is re-plotted in Fig.~\ref{fig:decoherence}(a) together with the result for setting (ii) in (b) for comparison. Because the logical encoding resides in the single-excitation subspace of each qubit pair, decoherence can cause the state to leak into the zero- and multi-excitation subspaces, which do not correspond to any valid promise answer. We therefore post-select on the single-excitation subspace of the eight physical qubits, retaining only measurement outcomes with one excited qubit and discarding all leakage events. 

True probabilities $P_{\rm true}(y)$ and post-selection success probabilities $P_{\rm succ}(y)$ are evaluated from the raw measurement outcomes as follows. Denote by $M = 8192$ the total number of measurement shots and by $M_{\rm se}$ the number of shots in which exactly one of the eight physical qubits is excited. The post-selection success probability is the fraction of shots surviving this filter, $P_{\rm succ}(y) = M_{\rm se}/M$. Within the retained single-excitation shots, $M_y$ further counts those in which the excitation resides in the physical qubits assigned to output port $y$, corresponding to the promise answer $y$, and the true probability is defined as $P_{\rm true}(y) = M_y/M_{\rm se}$. Thus $P_{\rm true}(y)$ characterizes the accuracy of the post-selected data, whereas the product $P_{\rm true}(y)P_{\rm succ}(y)$ gives the fraction of all shots that both survive post selection and return the correct promise answer. 

Table~\ref{tab:noisy} presents the simulation results of the HPP with random rotations for the two coherence time settings, averaged over 20 Haar-random basis rotations. For $(T_1,T_2)=(100,75)\,\mu\mathrm{s}$, both the true and post-selection success probabilities remain close to unity, showing that the advantage of indefinite causal order is robust against decoherence at the level of state-of-the-art superconducting processors. In contrast, for $(T_1,T_2)=(20,2)\,\mu\mathrm{s}$, comparable to the coherence times of the device used in this work, the true probabilities barely exceed the random-guessing value while most shots are discarded by post selection, indicating that the short coherence time severely limits the single-excitation encoding.

\section{\theappsec. Qutrit encoding approach}

In this section, we briefly discuss the qutrit encoding approach to address the lack-of-vacuum problem in the superconducting processors, as an alternative to the logical qubit encoding approach we adopt in the present work. The essence of the qutrit encoding approach is to expand the Hilbert space by utilizing the transmon's second excited state. For this, we denote the first three basis states for the superconducting qutrit by \(\{\ket{0}, \ket{1}, \ket{2}\}\), with \(\ket{0}\) being the ground state, and \(\ket{1}\) (\(\ket{2}\)) the first (second) excited states. A possible encoding is to encode a new qubit with two basis states $\ket{0}_Q$ and $\ket{1}_Q$ to the qutrit first and second excited states for the quantum switch experiment, i.e. \(\ket{0}_Q \coloneqq \ket{1}\) and \(\ket{1}_Q \coloneqq \ket{2}\), while treating the ground state \(\ket{0}\) as the independent vacuum state. The anharmonicity of the transmon qubits allows for selective manipulation of the excited states while leaving the ground state unchanged. Although single-qubit operations can be achieved in the excited subspace~\cite{liu2023performing}, it requires implementing the QHT in two subspaces of the same group of transmon qubits, i.e. \(\{\ket{0}_{Q,j}\}_{j=0}^{N-1}\) and \(\{\ket{1}_{Q,j}\}_{j=0}^{N-1}\), to achieve the coherent superposition of different paths. Despite targeting orthogonal excitation subspaces, the two QHTs must be implemented sequentially because the ground states of the qubits serve as shared resources for the single-qubit and CNOT gates in both circuits. Overall, this non-parallelism approximately doubles the circuit depth, making the qutrit encoding approach susceptible to decoherence and gate infidelity in superconducting platforms.

\begin{table}[t]
	\centering
	\caption{True probabilities $P_{\rm true}(y)$ and post-selection success probabilities $P_{\rm succ}(y)$ for two coherence time settings $(T_1,T_2)$, averaged over 20 random basis rotations. Uncertainties are the corresponding standard deviations.}
	\label{tab:noisy}
	\setlength{\tabcolsep}{9pt}
	\begin{tabular}{c cc cc}
		\hline \hline
		& \multicolumn{2}{c}{$(T_1,T_2)=(100,75)\,\mu\mathrm{s}$} & \multicolumn{2}{c}{$(T_1,T_2)=(20,2)\,\mu\mathrm{s}$} \\
		\cmidrule(lr){2-3}\cmidrule(lr){4-5}
		$y$ & $P_{\rm true}$ & $P_{\rm succ}$ & $P_{\rm true}$ & $P_{\rm succ}$ \\
		\midrule
		0 & $0.991\pm0.001$ & $0.934\pm0.001$ & $0.537\pm0.007$ & $0.232\pm0.004$ \\
		1 & $0.9902\pm0.0008$ & $0.9212\pm0.0009$ & $0.469\pm0.005$ & $0.199\pm0.003$ \\
		2 & $0.989\pm0.001$ & $0.923\pm0.001$ & $0.450\pm0.005$ & $0.198\pm0.003$ \\
		3 & $0.9889\pm0.0004$ & $0.921\pm0.003$ & $0.477\pm0.004$ & $0.194\pm0.003$ \\
		\hline \hline
	\end{tabular}
\end{table}

\end{document}